\documentstyle[editedvolume,numreferences,amssymb,bm,amsmath,graphicx]{crckapb} 

\begin{opening}
\title{Multiterminal counting statistics}

\author{Dmitri A. Bagrets, Yuli V. Nazarov}
\institute{Department of Applied Physics and \\
Delft Institute of Microelectronics and Submicrontechnology, \\
Delft University of Technology, Lorentzweg 1, 2628 CJ Delft, The Netherlands}

\end{opening}

\begin{document}



\begin{abstract}
\small 
  The review is given of the calculational schemes that allows for easy
evaluation of full current statistics (FCS) in multi-terminal mesoscopic systems. 
First, the scattering approach by Levitov {\it et.al } to FCS is outlined.
Then the multi-terminal FCS of the non-interacting electrons is considered.
We show, that this theory appears to be a circuit theory of $2\times 2$ matrices 
associated with Keldysh Green functions. Further on the FCS in the
opposite situation of mesoscopic systems placed in a strong Coulomb blockade limit
is discussed. We prove that the theory of FCS in this case turns out to be an elegant
extension of the master equation approach. We illustrate both methods
by applying them to the various three-terminal circuits. We study the FCS 
of electron transfer in the three-terminal chaotic quantum dot and compare it
with the statistics of charge transfer in the Coulomb blockade island with
three leads attached. We demonstrate that Coulomb interaction suppresses the relative 
probabilities of big current fluctuations. Finally, for the generic case of single 
resonance level the equivalence of scattering and master equation approach 
to FCS is established.

\end{abstract}

\begin{section}{Introduction}

The field of quantum noise in mesoscopic systems has been exploded
during the last decade, some achievements being summarized in a recent
review article. \cite{BlanterReview} 
While in classical systems shot noise is just a straightforward manifestation
of discreteness of the electron 
charge, it can be used in quantum system as unique tool to reveal
the information about the electron correlations and entanglement of various kinds. 
Measurement of fractional charge in Quantum Hall regime \cite{Frac_Charge}, noise 
measurements in atomic-size junctions \cite{Cron}, chaotic quantum 
dots\cite{Chaotic} and superconductors \cite{Kozhe,Jehl} are
milestones of the field of quantum noise. 
Starting from the pioneering work of B\"{u}ttiker\cite{Buttiker}, a special 
attention has been also paid to shot noise and noise correlations in the 
"multi-terminal" circuits. The correlations of currents flowing to/from different terminals 
can, for instance, reveal Fermi statistics of electrons. 
Such cross-correlations have been recently 
seen experimentally in Ref.~\cite{Tarucha,Oliver,Hall}. 

  A very important step in the field of quantum noise has been made by Levitov {\it et. al.} 
in \cite{LL,LLL} where an elegant
theory of {\it full counting statistics} (FCS) has been presented. This
theory provides an efficient way to investigate the current correlations
in the mesoscopic systems.
The FCS method concentrates on the evaluation of the probability distribution function
of the numbers of electrons transferred to the given terminals during 
the given period of time. 
It yields not only the noise, but all higher 
momenta of the charge transfer. 
The probability distribution also contains the fundamental 
information about big current fluctuations in the system.  

  Original FCS method~\cite{LL,LLL} was formulated for the scattering approach to 
mesoscopic transport. This made possible the study the
statistics of the transport through the disordered metallic conductor~\cite{Yakovets}
and the two-terminal chaotic cavity~\cite{BlanterSchomerus}. Muzykantskii and Khmelnitskii
have generalized the original approach by Levitov {\it et. al. }
to the case of the normal metal/su\-per\-con\-duc\-ting
contacts. The very recent application of the scattering approach
is the counting statistics of the charge pumping in the open quantum 
dots~\cite{Andreev,Levitov1,Mirlin}.

   However, scattering approach becomes unpractical in case of realistic layouts
where the scattering matrix is multi-channel, random and
cumbersome. For practical calculations one evaluates the FCS
with the semiclassical Keldysh Green function method~\cite{RefYuli} or with its 
reduction called the circuit theory of mesoscopic transport~\cite{General}. 
The Keldysh method to FCS was first proposed
by one of the authors. It has been applied to treat the effects of the weak localization 
corrections onto the FCS in the disordered metallic wires and later on to study  
the FCS in superconducting heterostructures~\cite{Belzig}.

   The above researches address the FCS of the non-interacting electrons in case
of  common two-terminal geometry. Although the cross-correlations for several 
multi-terminal layouts have been understood~\cite{BlanterReview}, 
the evaluation of FCS encountered difficulties. 
For instance, an attempt to build up FCS with "minimal correlation approach" 
\cite{BlanterSukhor} has lead to contradictions \cite{BlanterSchomerus}.
Very few is known about FCS of interacting electrons~\cite{Andreev1,Fazio}.
Since the interaction may bring correlations and entanglement of electron states 
the study of FCS of interacting electrons, particularly for the case of multi-terminal
geometry, is both challenging and interesting. 

   In the present work we review the calculational scheme that allows for easy
evaluation of FCS in multi-terminal mesoscopic systems. 
The paper is organized as follows. In the section II we outline the 
scattering approach by Levitov {\it et.al } to FCS. 
Section III is devoted to the multi-terminal FCS of the non-interacting electrons.
We show, that this theory appears to be a circuit theory of $2\times 2$ matrices 
associated with Keldysh Green functions~\cite{NazBag}. 
In the section IV we concentrate on FCS in the
opposite situation of mesoscopic systems placed in a strong Coulomb blockade limit. 
We prove that the FCS in this case turns out to be an elegant
extension of the master equation approach. We illustrate both methods
by applying them to the various three-terminal circuits. We study the FCS 
of electron transfer in the three-terminal chaotic quantum dot and compare it
with the statistics of charge transfer in the Coulomb blockade island with
three leads attached. We demonstrate that Coulomb interaction suppresses the relative 
probabilities of big current fluctuations. 
In section V we establish the equivalence of
scattering and master equation approach to FCS, by considering the statistics of
charge transfer through the single resonance level.  Finally, we summarize the results in 
the section VI.

\end{section}

\begin{section}{Scattering approach to FCS}

  In this section we review the current statistics of non-interacting electrons, using
the ideas of general scattering approach to mesoscopic transport. The problem of FCS in 
this framework has been solved about a decade ago by Levitov {\it et. al}~\cite{LL,LLL}.
Their work, concentrated on the two-terminal geometry
\cite{LLL}, became widely cited. But, unfortunately, their preceding paper\cite{LL}, devoted to 
multi-terminal systems as well, is only partially known. 
We will also demonstrate its relation to the current-current correlations 
in multi-terminal conductors, first investigated by M.B\"uttiker~\cite{Buttiker}. 

\begin{figure}[t]
\begin{center}
\includegraphics[scale=0.275]{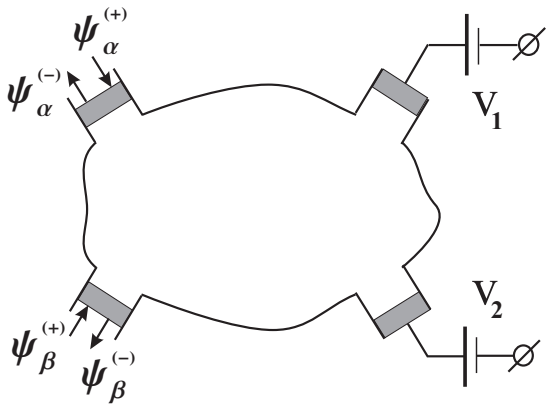}
\caption{ The "multi-terminal" mesoscopic conductor.  }
\label{Conductor}
\end{center}
\end{figure}

The most general mesoscopic system, eligible to scattering approach, is shown in 
Fig.~\ref{Conductor}. It is a phase coherent mesoscopic conductor, connected to 
macroscopic reservoirs
(leads, terminals) by means of $n$ external contacts ($n\geq 2$). 
We concern the situation when the system is placed in
the non-equilibrium condition, thus each terminal $\alpha$ is characterized by a
distribution function $f_\alpha(\varepsilon)$ of electrons over the energy. When
the terminal is in local equilibrium at electro-chemical potential $V_\alpha$ and 
a temperature $T_\alpha$, then  $f_\alpha(\varepsilon)$ reduces to Fermi distribution
$f_\alpha(\varepsilon) = \{1+\exp[(\varepsilon-eV_\alpha)/kT_\alpha]\}^{-1}$.

The conductor is a region of disordered or chaotic scattering.  
The scattering within conductor is assumed to be elastic. 
This is true at sufficiently low energies, such that
the energy dependent length of inelastic scattering  exceeds the sample size.
As to the terminals, it is conveniently to assume, that far from the contacts the longitudinal 
(along the lead) and transverse (across the lead)
motion of electrons are separable. Then at fixed energy $\varepsilon$ the transverse motion 
in the lead $\alpha$ is quantized and described by the integer index 
$n_\alpha=1 \dots M_\alpha(\varepsilon)$,
with $M_\alpha(\varepsilon)$ being the total number of propagating modes at given energy.
Each $n_\alpha$ corresponds to one incoming to the contact and one outgoing from the contact
$\alpha$ solution of the Shr\"odinger equation, which is usually referred as a 
scattering channel.

   The idea behind the scattering approach is that the electron transport can
be described as the one-particle scattering from the incoming channels to the 
outgoing channel in the leads. 
Very generally it can be described by a unitary scattering matrix $\hat S(\varepsilon)$. 
At given energy its dimension is equal to $M$ , $M=\sum_{\alpha} M_\alpha$ 
being the total number of transport channels. The energy dependence of the scattering matrix is
set by the "Thouless energy" $E_{\rm Th}=\hbar/\tau_d$, $\tau_d$ being
the traversal time through the system. At sufficiently low temperatures, voltages and 
frequencies,  such that $\max\{eV,kT,\hbar\omega\}\ll E_{\rm Th}$, one can 
conveniently disregard the energy
dependence and concentrate on the scattering matrix $\hat S(E_F)$, taken at Fermi surface.
Then the scattering approach gives the adequate description of the transport, provided the 
effects of Coulomb blockade can be neglected. This is justified, when the total conductance
$G$ of the sample is much larger that the conductance quantum $G_Q = e^2/2\pi\hbar$. 

  One starts by introducing the scattering matrix $\hat S$ of electrons 
at a narrow energy strip $\sim \max(eV,kT)\ll E_{\rm Th}$ in the vicinity of Fermi energy $E_F$.
The $S$-matrix relates the amplitudes $\psi^{(+)}_{m,\,\beta}$ of incoming electrons to the 
amplitudes $\psi^{(-)}_{n,\,\alpha}$ of the outgoing ones
(See Fig.~\ref{Conductor}): 
\begin{equation}
  \psi^{(-)}_{n,\,\alpha} = \sum_{m,\beta} S_{nm,\, \alpha\beta}
  \psi^{(+)}_{m,\,\beta}
\label{FullSMatrix}
\end{equation}
Here Latin indexes $(n,m)$ refer to the particular quantum scattering channel in given leads 
$(\alpha, \beta)$. (See e.g. \cite{BlanterReview,Buttiker,Beenakker} for a more detailed
introduction.) It is worth to mention that at this stage  $\psi^{(\pm)}_{n,\,\alpha}$
correspond to the amplitude of the wave function in the {\it exterior} area of the contacts.
Thus defined $S$-matrix incorporates the information about scattering in the
whole system. This includes the scattering from the disordered/chaotic regions of the 
conductor as well as multiple reflections from external contacts. 

   Remarkably, the FCS can be expressed in terms of the scattering matrix in 
quite a general way. It has been shown by Levitov {\it et. al.}~\cite{LL}.
The FCS approach deals with the probability
distribution $P(\{N_i\})$ for $N_i$ electrons to be transferred through the terminal
$i$ during time interval $t_0$. The detection time $t_0$ is assumed to be much
greater than $e/I$, $I$ being a typical current through the conductor.
This ensures that in average $\bar{N_i}\gg 1$. The probability 
distribution $P(\{N_i\})$ can be conveniently expressed via generating function ("action")
$S(\{\chi_i\})$ by means of Fourier transform
\begin{equation}
P(\{N_i\})= \int_{-\pi}^{\pi} \prod_i \frac{d \chi_i}{2 \pi} e^{-{\cal S}(\{\chi_i\}) -
i \sum_i N_i \chi_i}.
\label{Def}
\end{equation}
Here $\chi_i$ are the parameters of this generating function ("counting fields"), 
each of them being associated with a given terminal $i$. 
The generating function ${\cal S}(\{\chi_i\})$ 
contains the whole information about the irreducible moments of charge transfer
through the system, as well as the information about big current fluctuations. 
Indeed, it follows from Exp.~(\ref{Def}) that (higher-order) derivatives 
of ${\cal S}$ with respect to $\chi_i$, evaluated at $\chi_i =0$, give (higher-order) 
irreducible moments of $P(\{N_i\})$. First derivatives yield average currents 
to terminals, second derivatives correspond to the noises and noise correlations.

   The result of Levitov {\it et. al.} for the generating function ${\cal S}(\{\chi_i\})$
reads as follows. One considers the energy-dependent determinant
\begin{equation}
  {\cal S}_E(\{\chi_i\}) = {\rm ln}\,{\rm Det}\Bigl(1-\hat f_E + \hat f_E \hat S^\dagger \tilde 
S\Bigr)
\label{Det}
\end{equation}
where the matrix $(\hat f_E)_{mn,\,\alpha\beta} = f_\alpha (E)$ is
diagonal in the channel indexes $(m,n)$ and the matrix $\tilde S$ is defined as
${\tilde S}_{mn,\,\alpha\beta} = e^{i(\chi_\alpha-\chi_\beta)} S_{mn,\,\alpha\beta}$.
It has been proved that this expression represents a characteristic function of the
probability distribution of transmitted charge for electrons with energies in an
infinitesimally narrow energy strip in the vicinity of $E$. After that the complete
generating function ${\cal S}(\{\chi_i\})$  is just a sum over energies
\begin{equation}
 {\cal S}(\{\chi_i\}) = t_0\int {\cal S}_E(\{\chi_i\}) \frac{dE}{2\pi\hbar}  
\label{Sfull}
\end{equation}
This expression reflects the fact that electrons at different energies are transferred
independently, without interference. Therefore they yield additive independent contributions
to the generating function. 

   The general results for the FCS, contained in Eqs.~(\ref{Det}) and (\ref{Sfull}), 
provides the elegant way to derive the pioneering results of M. B\"uttiker
\cite{Buttiker},
concerning the current-current correlations in the multiterminal conductors. 
One considers the correlation function
\begin{equation}
  P_{\alpha\beta}(t-t') = \frac{1}{2}\left\langle 
\Delta\hat I_\alpha(t)\Delta\hat I_\beta(t') + \Delta\hat I_\beta(t')\Delta\hat I_\alpha(t) 
\right\rangle
\label{Ps}
\end{equation}
of currents fluctuations in contacts $\alpha$ and $\beta$. The fluctuation
$\Delta\hat I_\alpha(t)$ is defined as 
$\Delta\hat I_\alpha(t) = \hat I_\alpha(t)- \langle I_\alpha \rangle$, with 
$\hat I_\alpha(t)$ being the current operator in the lead $\alpha$ and
$\langle I_\alpha \rangle$ being its mean steady value under given non-equilibrium
conditions. The Fourier transform $P_{\alpha\beta}(\omega)$ of current-current
correlations functions is sometimes referred to as noise power. 
Throughout this article we concentrate on zero-frequency (shot-noise) limit
of current correlations $P_{\alpha\beta} \equiv  P_{\alpha\beta}(\omega=0)$. 
At low frequencies both $\langle I_\alpha \rangle$  and $P_{\alpha\beta}$ are readily
expressed via the first (second) moments of number of transferred electrons through 
the corresponding leads
\begin{equation}
 \langle I_\alpha \rangle  = \frac{e}{t_0} \langle N_\alpha \rangle, \qquad
 P_{\alpha\beta} = \frac{e^2}{t_0} \langle \Delta N_\alpha \Delta N_\beta \rangle\
\end{equation}
Therefore the second derivative 
$\frac{\partial^2}{\partial \chi_\alpha \partial\chi_\beta} {\cal S}(\{\chi_i\})$ 
of the action yields the result for the 
shot-noise correlation function $P_{\alpha\beta}$. It can be reduced to the
following  form
\begin{equation}
P_{\alpha\beta} = \frac{e^2}{\pi\hbar} \int\,dE \sum_{\gamma\mu}{\rm Tr}
\left\{ A_{\gamma\mu}^{\alpha} A_{\mu\gamma}^{\beta} \right\}
f_\gamma(E)(1-f_\mu(E))
\label{MultiCorr} 
\end{equation}
Here the matrix $A_{\gamma\mu}^{\alpha}$ is defined via $S$-matrix as
\begin{equation}
 A_{\gamma\mu}^{\alpha}(E) = \delta_{\alpha\gamma}\delta_{\alpha\mu} - 
  S^\dagger_{\alpha\gamma}(E) S_{\alpha\mu}(E)
\end{equation}
This is precisely the result obtained by M.~B\"uttiker
prior the development of FCS approach.

  Up to the moment the scattering matrix $\hat S$ was not specified. 
In order to evaluate the FCS, one needs to construct such a matrix, so that
to take into account the scattering properties of the concrete mesoscopic system
at hand. To accomplish this task, one may proceed as follows.
Very generally, one can separate a mesoscopic layout into primitive
elements: nodes and connectors. The nodes are similar to the terminals, in the
sense, that each of them can be characterized by some non-equilibrium isotropic distribution
function of electrons over the energies. The illustrative example of the node is
chaotic quantum dot.  The scattering within the node is assumed to be random (chaotic). 
The connectors represent themselves either internal or external
contacts in the systems, scattering properties of which are supposed to be known.
The actual implementation of connectors can be quite different:
quantum point contacts, tunnel junctions, diffusive wires, etc.   The purpose of
the above separation is to ascribe to each node and to connector the particular 
unitary scattering matrix with known properties.
Then the scattering matrix $\hat S$ of the entire system
can be unambiguously expressed via these primitive scattering matrices.

\begin{figure}[t]
\begin{center}
\includegraphics[scale=0.25]{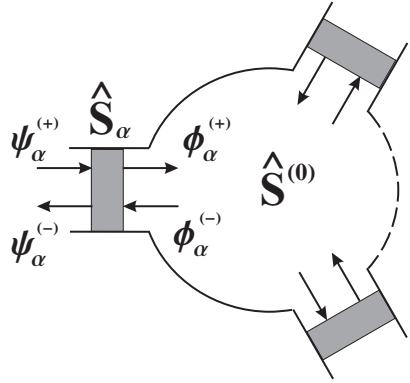}
\caption{The building blocks of a scattering matrix of the multi-terminal
chaotic quantum dot. The matrix $\hat S_\alpha$ 
describes the scattering due to the contact $\alpha$. The matrix $\hat S^{(0)}$ describes
the chaotic scattering within the dot.
\label{ChaoticDot} }
\end{center}
\end{figure}

   We outline this scheme for the simple system, consisting of the single 
node only. A good example of such system is 
a multi-terminal chaotic quantum dot, shown in Fig.~\ref{ChaoticDot}.  
It is coupled to $n$ external
reservoirs via $n$ contacts.  Each contact $\alpha$ is described by a unitary
scattering matrix $\hat S_\alpha$. (Its choice depends on the experimental
realization of this given contact.) Let $\phi^{(\pm)}_{m,\,\alpha}$ be the
coefficients of electron wave function, taken at the {\it interior} boundary of the 
contact. The amplitudes $\phi^{(+)}_{m,\,\alpha}$ and $\phi^{(-)}_{m,\,\alpha}$ 
correspond to the outgoing (incoming) waves from (to) the the contact $\alpha$
respectively, as shown in Fig.~2. The scattering matrix $\hat S_\alpha$ relates
the vector $c^{\rm in} = (\psi^{(+)}_{\alpha}, \phi^{(-)}_{\alpha} )^T$  of the
incident electron waves to the vector 
$c^{\rm out} = (\psi^{(-)}_{\alpha}, \phi^{(+)}_{\alpha} )^T$ of the outgoing waves:
\begin{equation}
 c^{\rm out} = \hat S_\alpha \,c^{\rm in} = \left(
\begin{array}{cc}
r_\alpha & t'_\alpha \\
t_\alpha & r'_\alpha \\
\end{array}\right) c^{\rm in}
\label{ConnectorsS}
\end{equation}
Here the square blocks $r_\alpha$ ($r'_\alpha$)  of the size $(M_\alpha\times M_\alpha)$ 
describe the electron reflection back to the reservoirs or back to the dot, respectively.  
The off-diagonal blocks $t_\alpha$ ($t'_\alpha$) of the same dimension 
are responsible for the transmission through the contact $\alpha$.

  Similarly, the chaotic scattering of electrons inside the dots is described by the 
scattering matrix $\hat S^{(0)}$. By analogy to Eq.~(\ref{FullSMatrix}), 
it relates within the dot amplitudes $\phi^{(+)}_{m,\,\beta}$ to
$\phi^{(-)}_{n,\,\alpha}$, i.e.
\begin{equation}
  \phi^{(-)}_{n,\,\alpha} = \sum_{m,\beta} 
  S^{(0)}_{nm,\, \alpha\beta}
  \phi^{(+)}_{m,\,\beta}
\label{DotS}
\end{equation}
The dimension $(M\times M)$ of $\hat S^{(0)}$ is equal to that of the matrix $\hat S$.

   It is possible now to express the scattering matrix $\hat S$ of the entire sample in
terms of $\hat S_\alpha$ ($\alpha=1\dots n$) and $\hat S^{(0)}$. One can use the set
of equation~(\ref{ConnectorsS}) and (\ref{DotS}) in order to find the linear relations
between amplitudes $\psi^{(+)}$ and $\psi^{(-)}$. According to Eq.~(\ref{FullSMatrix}) 
it will uniquely determine the matrix elements of $\hat S_\alpha$. The result reads:
\begin{equation}
 \hat S = \hat r + \hat t'\hat S^{(0)}\frac{1}{1-\hat r' \hat S^{(0)}}
 \hat t
\label{EntireS} 
\end{equation}
Here $\hat r$ ($\hat r'$) and $\hat t$ ($\hat t'$) are block diagonal matrices
with elements $\hat r_{mn,\,\alpha\beta} = \delta_{\alpha\beta}(\hat r_\alpha)_{mn}$
and $\hat t_{mn,\,\alpha\beta} = \delta_{\alpha\beta}(\hat t_\alpha)_{mn}$. 
The denominator of Eq.~(\ref{EntireS}) describes the multiple reflection 
due to contacts within the dot. It can be verified, that $\hat S$ is unitary, 
$\hat S^\dagger \hat S =1$, provided $\hat S^{(0)}$ and $\hat S_\alpha$ are also unitary.
The analogous construction of $S$-matrix can be 
in principle implemented for the more complicated layouts.

    To evaluate FCS one substitutes $\hat S$ into Eqs.~(\ref{Det}) and (\ref{Sfull}) in the
form (\ref{EntireS}). The matrices $\hat S_\alpha$ are fixed by the choice of the
properties of the contacts. However, the matrix $\hat S^{(0)}$ is random describing
chaotic scattering inside the dot. This means that the answer should be {\it averaged}
over all possible realization of $\hat S^{(0)}$. A reasonable assumption is that
$\hat S^{(0)}$ is uniformly distributed in the space of all unitary matrices~\cite{Beenakker}. 

   Thereby one solves the problem for a single node. For less
trivial system, comprising two and more nodes, the expression for the $\hat S$-matrix,
corresponding to Eq.~(\ref{EntireS}), becomes more involved. 
Moreover one has to average over unitary matrices separately for each node.
Besides, to describe a simple system of diffusive wire one has eventually to go to
the limit of infinitely many nodes.  All this makes the scattering approach extremely 
inconvenient. Fortunately, one can treat these problems with a 
more flexible semiclassical Green function method, which we outline in the next
section. It is applicable, since we assume that the conductance of the
system $G\gg G_Q$.

\end{section}

\begin{section}{Circuit theory of FCS in multi-terminal circuits}

  In this section we formulate the semiclassical theory of FCS of 
non-interacting electrons in case
of multi-terminal geometry. Our approach is based on the Keldysh 
Green function method. We develop the scheme to evaluate the FCS in the arbitrary
multi-terminal mesoscopic system. For that we will use the circuit theory of mesoscopic 
transport~\cite{General}. Next, we illustrate this scheme, considering the big 
fluctuations of current in the three-terminal chaotic quantum dot. 
In the end of the section we discuss the shot-noise correlations
and give the convenient expressions that depend on the scattering properties 
of connectors only, and do not involve the scattering inside
the cavity.  

\begin{subsection}{Circuit theory approach to the FCS.}
We start by introducing current
operators $\hat I_i$, each being associated with the current to/from a certain 
terminal $i$. Extending the method of \cite{RefYuli} we introduce a 
Keldysh-type Green function defined by
\begin{equation}
\Big( i\frac{\partial}{\partial t} - \hat
H - \frac12 \bar\tau_3 \sum_i \chi_i(t)\hat I_i \Big) \otimes \check G(t,t')= 
\delta(1-1')
\label{green}
\end{equation} 
Here we follow notations of a comprehensive review \cite{Rammer}: 
$\chi_i$ are time-dependent parameters, $\bar \tau_3$ is a
$2 \times 2$ matrix in Keldysh space, $\hat H$ is the one-particle Hamiltonian 
that incorporates all information about the system layout, including boundaries, 
defects and all kinds of elastic scattering. We use "hat", "bar" and "check" to 
denote operators in coordinate space, matrices in Keldysh space
and operators in direct product of these spaces respectively. 
The current operator in Eq.~(\ref{green}), associated with the terminal $i$,
is defined as 
$\hat I_i = \int d^3 x \Psi^\dagger({\bf p}/m)\Psi\, {\bf \nabla} F_i(x)$.
Here $\Psi$ is the usual Fermi field operator and ${\bf \nabla} F_i(x)$ is
chosen such a way that the spatial integration is restricted to the cross-section
and yields the total current through the given lead.
The Eq. (\ref{green})
defines the Green function unambiguously provided boundary conditions
are satisfied: $\check G(t,t') \equiv \bar G(x,x';t,t')$ approaches the common 
equilibrium Keldysh Green functions
$\check G^{(0)}_i(t-t')$ provided $x,x'$ are sufficiently far in the terminal 
$i$.
These $\check G^{(0)}_i(t-t')$ incorporate information
about the state of the terminals: their voltages $V_i$ and temperatures $T_i$.

  In the following we will operate with the cumulant generating function ("action")
$S(\{\chi_i\})$ defined as a sum of all closed diagrams 
\begin{equation}
 e^{-S(\{\chi_i\})} = \left\langle 
T_\tau e^{i \sum_{i=1}^N 
\int\limits_{-\infty}^{+\infty} d\tau \chi_i(\tau)\check I^{(i)}(\tau)}
{\widetilde T}_\tau e^{ i \sum_{i=1}^N 
\int\limits_{-\infty}^{+\infty} d\tau \chi_i(\tau)\check I^{(i)}(\tau)}
\right\rangle
\label{QM_Action}
\end{equation}
Here $T_\tau$ ($\widetilde T_\tau$) denotes the (anti) time ordering operator.
One can see by traditional diagrammatic methods \cite{Rammer} that the 
expansion of $S(\{\chi_i\})$ in powers of $\chi_i(t)$ generates all possible 
irreducible diagrams for
higher order correlators of $\hat I_i(t)$ and thereby incorporates all the 
information about statistics of charge transfer. 
If we limit our attention 
to low-frequency limit of current correlations, we can keep time-independent 
$\chi_i$. In this case, the Green functions are functions of time difference 
only and the Eq.(\ref{green}) separates in energy representation. Then the
action $S(\{\chi_i\})$ can be conveniently expressed via the average
$\chi$-dependent currents $I_i(\{\chi_i\})$ in the following way
\begin{equation}
\frac{i}{t_0}\frac{\partial S}{\partial \chi_i} = I_i(\{\chi_i\}) \equiv
\int \frac{d\varepsilon}{2\pi}{\rm Tr} \ {\bf(} \bar \tau_3 \hat I_i  
\check 
G(\varepsilon) {\bf)}
\label{action}
\end{equation}
where $t_0$ denotes the time of measurement. 
Thus defined cumulant generating functions allows to evaluate
the probability for $N_i$ electrons to be transferred 
to the terminal $i$ during time interval $t_0$ in accordance with the general 
relation (\ref{Def}).


  Up to the moment the above Eqs. (\ref{green}) and (\ref{QM_Action})
can be used to define the statistics of any quantum mechanical variable. 
However we are interested in the statistics of the {\it charge} transfer.
Since the charge is the conserved quantity, the proper construction of
current operators $\hat I_i$ requires the  gauge invariance of the Hamiltonian.
Therefore fully quantum mechanical scheme includes the "counting" fields $\chi_i(t)$
as {\it gauge} fields. (See the contribution of L.Levitov in this book.)
It means, that the initial physical Hamiltonian $H({\bf q},{\bf p})$ of the
system should be replaced by the $\chi$-dependent Hamiltonian
$H_{\chi} = H({\bf q},{\bf p} - \frac{1}{2}\bar\tau_3\sum_i \chi_i{\bf\nabla} F_i)$.
Thus the appearance of the "counting" fields in the problem is similar
to the inclusion of a vector potential ${\bf A}(\bf r)$ of an ordinary electromagnetic
gauge field. The crucial difference is the change of sign of interaction at the
forward and backward branches of the Keldysh contour, that is reflected by the
presence of $\bar \tau_3$ matrix. Then the $\chi$-dependent Green function
obeys the equation of motion, written with the use of $\chi$-dependent Hamiltonian
$H_\chi$. Accordingly, the "action" $S(\{\chi_i\})$ is defined as a sum of
all closed diagrams with respect to the interaction $H_{\rm int} = H_{\chi} - H$.

Thus defined
constructions are locally gauge invariant with respect to rotation in the Keldysh
space. In particular, the average current (\ref{action}) is a conserved quantity,
provided the current operator is defined as 
$\hat I_i = \partial H_{\chi}(t)/{\partial \chi_i}$. 
Therefore it is possible to perform the local gauge transformation 
$\psi ' = \exp\{-(i/2)\bar\tau_3\sum_i \chi_i F_i\}\,\psi$
in order to eliminate the $\chi$-dependent terms from the
equation of motion(\ref{green})~\cite{RefYuli,Belzig}.
The $\chi$-dependence of $\check G$ is thereby transferred to the boundary conditions:
the gauged Green function far in each terminal shall approach 
$\check G_i(\epsilon)$ defined as
\begin{equation}
\check G_i(\epsilon) = \exp (i \chi_i \bar \tau_3/2) \check G^{(0)}_i(\epsilon) 
\exp (-i \chi_i \bar \tau_3/2)
\label{boundary}
\end{equation}
Here $\check G^{(0)}_i(\epsilon)$ corresponds to the equilibrium Keldysh Green 
function sufficiently far in the given terminal $i$.
The precise form of the Green functions $\check G^{(0)}_i(\epsilon)$ will be explicitly
given below in the text. [See Eq. (\ref{G0})]

  Let us also note that the modification of current operators $\hat I_i$ by the
"counting" fields can be safely omitted at energies near the Fermi surface, where one can
linearize the electron spectrum.  This is possible in the semiclassical 
approximation when the variation of ${\bf\nabla} F_i$ 
is small at the scale of the Fermi wave length. Expanding 
the $H_{\chi}$ on the Fermi shell, one arrives to Eqs.~(\ref{green}) and (\ref{QM_Action}).
Afterwards it is feasible to verify the result (\ref{boundary}), with making use of
the semiclassical Eilenberger equations. (For the details we refer the reader to
the paper of W.Belzig in this book.)
  
In the present form, the Eq.~(\ref{green}) with relations 
(\ref{boundary},\ref{action}) solves the problem of determination of the 
FCS for any arbitrary system layout: one just has to find exact 
quantum-mechanical solution of a Green function 
problem. This is hardly constructive, and we proceed further by deriving a
simplified semiclassical approach. First, we note that even in its exact 
quantum-mechanical form the Eq.(\ref{green}) possesses an important property. We consider 
the quantity  defined similar to standard definition of current density,
$\bar j^{\alpha}(x,\epsilon) \equiv \lim_{x \rightarrow x'}(\nabla'^{\alpha}-
\nabla^{\alpha}) \bar G(x,x';\epsilon)/m$.
By virtue of Eq.(\ref{green}) this quantity conserves so that
\begin{equation}
{\partial \bar j^{\alpha}(x,\epsilon)}/{\partial x^{\alpha}} =0
\label{conservation}
\end{equation}
This relation looks like the conservation law of particle current at a given
energy. However, this relation contains more information since it is a 
conservation law for a $2\times2$ {\it matrix} current.

Next, we construct a theory which makes use of this
conservation law. We concentrate on the semiclassical Green function in coinciding points,
$\bar G(x,\epsilon)\equiv i\bar G(x,x';\epsilon)/\pi\nu$, where $\nu$ is a density
of states at Fermi energy.  So defined Green function has been introduced in several 
semiclassical theories. \cite{Rammer,Larkin,Noneq} 
It satisfies the normalization condition 
$\bar G^2=\bar 1$. We relate the "current density" $\bar j$ to gradients and/or
changes of $\bar G(x)$, very much like the electric current density
is related to the voltage in circuit theory of electric conductance.
Following the approach of the circuit theory~\cite{General},
we separate a mesoscopic layout
into elements: nodes and connectors, so that the $\bar G(x)$ is constant across 
the nodes and drops across the connectors. 
One may associate a graph with each 
circuit, so that its lines $(i,j)$ would denote the connectors, and internal 
and external vertices correspond to the nodes and terminals, respectively. 
(See Fig.~\ref{CircuitG}) 
This separation of actual layout is rather heuristic, similar to
separation of an electric conductor of a complicated geometry onto nodes
and circuit theory elements. The bigger the number and the finer the
mesh of the nodes and connectors, the better the circuit theory
approximates the actual layout.  The idea for this separation is completely analogous to the one
considered in the section 2.  The nodes are similar to the terminals,
the difference is that $\bar G$ is fixed in the terminals and yet to be
determined in the nodes. The $\bar G$ in nodes are determined from Kirchoff 
rules reflecting the conservation law (\ref{conservation}): sum of the matrix 
currents from the node over all connectors should equal zero at each energy. 
For this, we should be able to express the matrix current via each connector 
as a function of two  matrices $\bar G_{i,j}$ at its ends.

\begin{figure}[t]
\begin{center}
\includegraphics[scale=0.3]{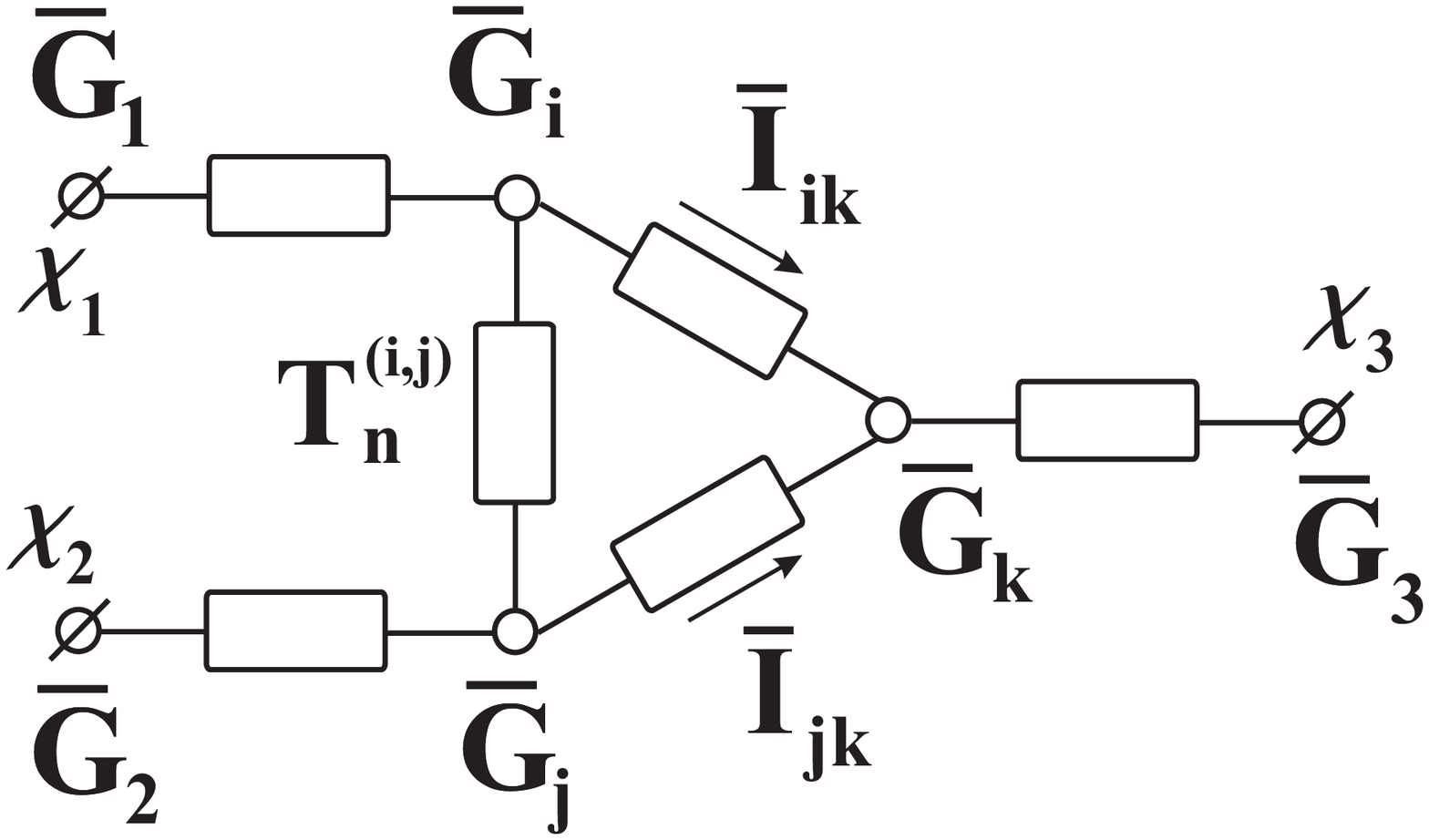}
\caption{The graph of the circuit theory, associated with a 3 terminal
mesoscopic system. 
$\bar G_1$, $\bar G_2$, $\bar G_3$ in the terminal 
are fixed by the boundary condition~(\ref{boundary}). $T_{n}^{ij}$ 
define the transport properties of a connector. $\bar I_{ik}$ and 
$\bar I_{jk}$ denote the currents, flowing from the nodes $i$ and $j$ 
into the node $k$. $\chi_1$, $\chi_2$, $\chi_3$  are counting fields
in the terminals.
\label{CircuitG} }
\end{center}
\end{figure}

   To accomplish this task, consider the connector $(i,j)$,  linking to nodes $i$ and $j$,
It can be quite generally characterized by a set of transmission eigenvalues 
$T_n^{(ij)}$\cite{General,Noneq}. 
The problem to solve is to express matrix current $\bar{I}_{ij}$ via the connector in terms of 
$\bar{G}_{i(j)}$. This problem shall be addressed by a more microscopic approach 
and was solved in \cite{Noneq} for Keldysh-Nambu matrix structure of $\check G$. 
It is a good news that the derivation made in \cite{Noneq} does not depend on 
concrete matrix structure and can be used for the present problem without any 
modification yielding
\begin{equation}
\label{current}
  \bar{I}_{ij} =\frac{1}{2\pi}\sum_n\int dE 
  \frac{T^{(ij)}_n\left[\bar G_i,\bar G_j\right]}{
    4+T^{(ij)}_n\left(\{\bar G_i, \bar G_j\} - 2 \right)}\,.
\end{equation}
Each connector $(i,j)$ in the layout contributes to the total $\chi_i$-dependent 
action (\ref{action}). The corresponding $S_{ij}$ contribution 
to the action should be found from the relation (\ref{action}) and
reads:\cite{Belzig}
\begin{equation}
  S_{ij}(\chi)=\frac{-t_0}{2\pi}\sum_n\int dE\, {\rm Tr}
  \ln\left[1+\frac{1}{4} T^{(ij)}_n\left(\{\bar G_i, \bar G_j\}-
2\right)\right]\;.
\label{action-connector}
\end{equation}

Now we are ready to present a set of circuit theory rules that enables
us to evaluate the FCS for an arbitrary mesoscopic layout. 
 
(i) The layout is separated onto terminals, nodes, and connectors.

(ii) The $\bar G_{j}$ in each terminal $j$ is fixed by relation (\ref{boundary}) 
thus incorporating information about voltage, temperature and counting field 
$\chi$ in each node. 

(iii) For each node $k$, the matrix current conservation yields a Kirchoff equation 
$\sum_i \bar I_{ik}=0$, where the summation is going over all connectors $(i,k)$ 
attached to node $k$, and $\bar I_{ik}$ are expressed with
(\ref{current}) in terms of $\bar G_{i(k)}$.

(iv) The solution of resulting equations with condition $\bar G_{k}^2=1$ fixes 
$\bar G_{k}$ in each node.

(v) The total action $S(\chi)$ is obtained by summing up the contributions 
$S_{ij}(\{\chi_i\})$ of individual connectors, those are given by 
(\ref{action-connector}):
$
S(\{\chi_i\})=\sum_{(i,j)} S_{ij}(\{\chi_i\})
$

(vi) The statistics of electron transfer is obtained from the  
relation (\ref{Def}).

 In the end of this subsection we discuss the limits of applicability 
of the whole scheme. By virtue 
of semiclassical approach, the mesoscopic fluctuations coming from interference 
of electrons penetrating different connectors are disregarded. So that, we 
assume that conductivities of all connectors are much bigger than conductance 
quantum $e^2/\pi\hbar$. The same condition provides the absence of 
Coulomb blockade effects in the system. Besides, we have disregarded the 
possible processes of {\it inelastic relaxation} in the system. The latter can be 
eventually taken into account, since the use of Keldysh Green functions technique 
allows for perturbation treatment of interaction and relaxation. However, it would 
considerably complicate the scheme. The point is that the inelastic 
scattering would mix up the $\bar G(\varepsilon)$ at different energies, so that one can not 
solve the circuit theory equations separately at each energy.

\end{subsection}

\begin{subsection}{The statistics of charge transfer in chaotic quantum dots}

As an illustration of the above scheme, we will consider in the second part of this 
section the FCS of the 3-terminal chaotic quantum dot. The system is sketched 
in the inset of Fig.~2. The heuristic circuit, associated with
this mesoscopic system is shown by dashed lines. It includes only 3 terminals, 
3 arbitrary connectors, associated with  
the contacts, and the node $\{4\}$, representing the quantum dot itself.
This separation is valid provided the cavity is in the {\it quantum}
chaotic regime. (See~\cite{Aleiner} for definition). This regime corresponds to 
full isotropization of the Green function $\check G(x,x',\epsilon)$ within the dot,
so that $\bar G_4(\varepsilon)$ can be regarded as a constant at a given energy.

Since the normalization $\bar G_k^2=1$ holds for each vertex, we
use the parametrization $\bar G_k={\bf g}_k \cdot \bm{\tau}$,
${\bf g}_k \cdot {\bf g}_k =1$.  Here ${\bf g}_k$ is a 3-D vector, and
$\bm{\tau} = (\bar\tau_1,\bar\tau_2,\bar\tau_3)$. With the use of this parametrization
the anticommutator $\{\bar G_i, \bar G_k\}$ is proportional to the unity matrix and takes the form 
of scalar product $\frac{1}{2}\{\bar G_i, \bar G_k\} = {\bf g}_i \cdot {\bf g}_k$.
In the absence of counting fields
the Green functions in the terminals  corresponds to the equilibrium Keldysh Green 
function~\cite{Rammer}
\begin{equation}
\bar G_k^{(0)} = \left(
\begin{array}{cc}
 1-2f_k & -2 f_k \\
 -2(1-f_k) & 2f_k-1
\end{array} \right), 
\label{G0}
\end{equation}
where Fermi distribution function $f_k(E)=\{ \exp[(E-eV_k)/T_k] + 1\}^{-1}$
accounts for the bias voltages $V_k$ and the temperatures $T_k$ in the 
terminals.
The $\chi_i$-dependence of $\bar G_k(\chi)$ is then given by Eq.~(\ref{boundary}). 

We see that Green function $\bar G_4(\chi)={\bf g}_4 \cdot \bm{\tau}$ in the dot
is in fact the only function to find. 
For that, we proceed by applying the current conservation law,
$\sum_{k=1}^3 \bar I_{k,4}=0$, inside the dot. We present 
the currents $\bar I_{k,4}$ given by~(\ref{current}) in the form
\begin{equation}
\bar I_{k,4}=\frac{1}{2}Z_k({\bf g}_k \cdot {\bf g}_4)
[\bar G_k,\bar G_4],
\label{dotcurrents}
\end{equation}
where the scalar function $Z_k(x)$ incorporates the information about transmission eigenvalues
in each connector $k$:
\begin{equation}
Z_k(x)\equiv \sum\limits_n T_n^{(k,4)}/[2+T_n^{(k,4)}(x-1)].
\label{Zfunction}
\end{equation}
It can be evaluated for any particular distribution 
$\rho(T)$ of transmission eigenvalues in the given connector and 
completely  defines its scattering properties.
For a example, if we denote by $R_{\rm Q}=\pi\hbar/e^2$ the resistance quantum, then 
$R^{-1}_k=2 R_{\rm Q}^{-1} Z_k(1)$ is an inverse 
resistance of the connector. One can also express the Fano factor
$F_k=\langle T(1-T) \rangle/\langle T\rangle$, associated with the given connector as 
$F=1-2 (d/dx) {\log Z_k(x)\bigr |}_{x=1}$.

With the use of $Z_k(x)$ the conservation law 
can be efficiently rewritten as 
$[\sum\limits_{k=1}^{3} p^k \bar G_k, \bar G_4]=0$,
where $p^k=Z_k\left({\bf g}_k \cdot {\bf g}_4 \right)$. This relation suggests
to look for the vector ${\bf g}_4$ in the form 
$
{\bf g}_4 = M^{-1}\sum\limits_{k=1}^3\,p^k\, {\bf g}_k,  
$
with $M(\chi)$ being an unknown normalization constant.
Using the normalization condition \\
${\bf g}_4 \cdot {\bf g}_4=1$ we obtain the
following set of equations 
\begin{equation}
\label{mapping}
p^i  =  Z_i\Bigl(M^{-1}\sum\limits_{j=1}^3 g_{ij}(\chi)\,p^j\Bigr), \quad
M^2  =  \sum\limits_{i,j=1}^3 g_{ij}(\chi) p^i p^j
\end{equation}
where the scalar product $g_{ij}(\chi)={\bf g}_i(\chi) \cdot {\bf g}_j(\chi)$ 
between terminal Green function 
is expressed in terms of Fermi distributions as follows
\begin{eqnarray}
g_{ij}(\chi) 
= (1-2f_i)(1-2f_j) 
+ 2\,e^{i(\chi_i-\chi_j)} f_i(1-f_j) + 2\,e^{-i(\chi_i-\chi_j)} f_j(1-f_i) 
\nonumber
\end{eqnarray}
The Green function $\bar G_4$ then is found from the solution $\{p^i(\chi), M(\chi)\}$
of this set of equations.  In the general situation the function $Z_k(x)$ takes 
the form~(\ref{Zfunction}) and therefore Eq.~(\ref{mapping}) represents the set 
of non-linear equations. However, their solution can be relatively easy 
found numerically using the method of subsequent iterations.

The total action can be found by applying the rule (v) of circuit theory
and reads
\begin{equation}
S(\chi) = \frac{\displaystyle t_0}{\displaystyle \pi} 
\sum\limits_{i=1}^3 \int d\epsilon\,
S_i\bigl( {\bf g}_i \cdot {\bf g}_4  \bigr)
\label{FCS-Action}
\end{equation}
where
\begin{equation}
{\bf g}_i \cdot {\bf g}_4 = M^{-1}(\chi)\sum\limits_{j=1}^3 g_{ij}(\chi)\,p^i(\chi) 
\end{equation}
Here partial contributions $S_k(x)$ from each connector
in Eq. (\ref{FCS-Action}) has to be determined from
the relation $\frac{\partial}{\partial x} S_k(x)=-Z_k(x)$, provided $S_k(1)=0$.
It follows from the Eqs.~(\ref{action}), (\ref{dotcurrents}) and 
normalization condition ${\bf g}_4 \cdot {\bf g}_4 = 1$.

  We consider three particular types of connectors: 
tunnel(T), ballistic(B) and diffusive(D). 
Their corresponding contributions to action read as~\cite{RefYuli}: 
\begin{eqnarray}
S_T(x) &=& -\frac{1}{2}(R_0/R) (x-1), \\
S_B(x) &=& - (R_0/R) \log[(1+x)/2], \\
S_D(x) &=& -\frac{1}{4} (R_0/R) \log^2(x+\sqrt{x^2-1})
\end{eqnarray}
with $R$ being a resistance of the connector.
The tunnel connector represents the tunnel junction, so that  $T_n\ll 1$ for all $n$.
Ballistic connector corresponds to the quantum point contact, with  $N$ open channels.
The last expression comes from universal transmission distribution
$\rho(T)= R_0/2RT\sqrt{1-T}$ for any diffusive contact.

Analytical results for FCS~(\ref{FCS-Action}) can be readily obtained only for the system 
with tunnel connectors. To assess general situation we found ${\bf g}_4$
for given $\chi_i$ numerically.
To find the probability distribution, we evaluated the
integral~(\ref{Def}) in the saddle point approximation, assuming
$\chi_i$ has to be complex numbers. Saddle point approximation is always valid
in the low frequency limit we consider, since in this case both action
$S$ and number of transmitted particles $N_i=I_i t_0/e \gg 1$. 
Due to the current conservation law 
only two of three counting fields ${\chi_i}$ are independent, and one can set 
$\chi_3=0$. The relevant saddle point of the function 
$\Omega(\chi) = S(\chi)+ i\chi_1 I_1 t_0/e + i\chi_2 I_2 t_0/e$ always 
corresponds to purely imaginary numbers $\{\chi_1^*,\chi_2^*\}$.
The probability reads $P(I_1,I_2)\approx \exp[-\Omega(\chi^*)]$. Evidently, 
$\Omega(\chi^*)$
is the Legendre transform of the action, and it can be regarded
as implicit function on $I(\chi^*)$.

\begin{figure}[t]
\begin{center}
\includegraphics[scale=0.45]{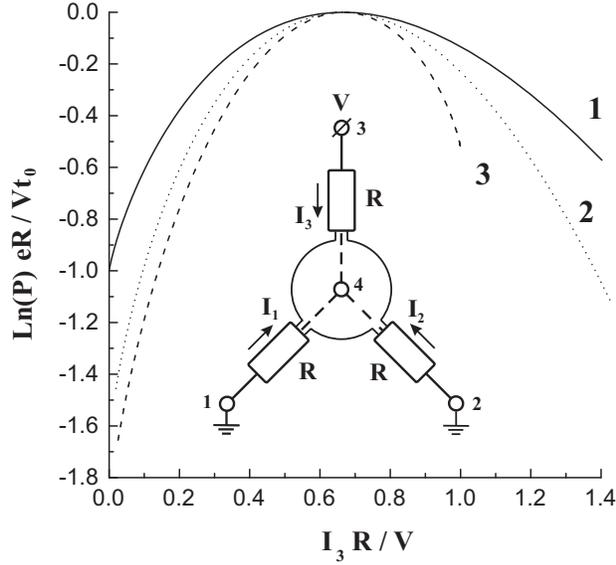}
\caption{The logarithm of the current probabilities in the 3-terminal 
chaotic quantum dot as a function of $I_3$, under condition $I_1=I_2$. 
The insert presents the system configuration. The resistances $R$
of all connectors are assumed to be equal. 1 - tunnel connectors,
2 - diffusive connectors, 3 - ballistic connectors. 
\label{LogPCh} }
\end{center}
\end{figure}

\begin{figure*}
\begin{center}
\includegraphics[scale=0.6]{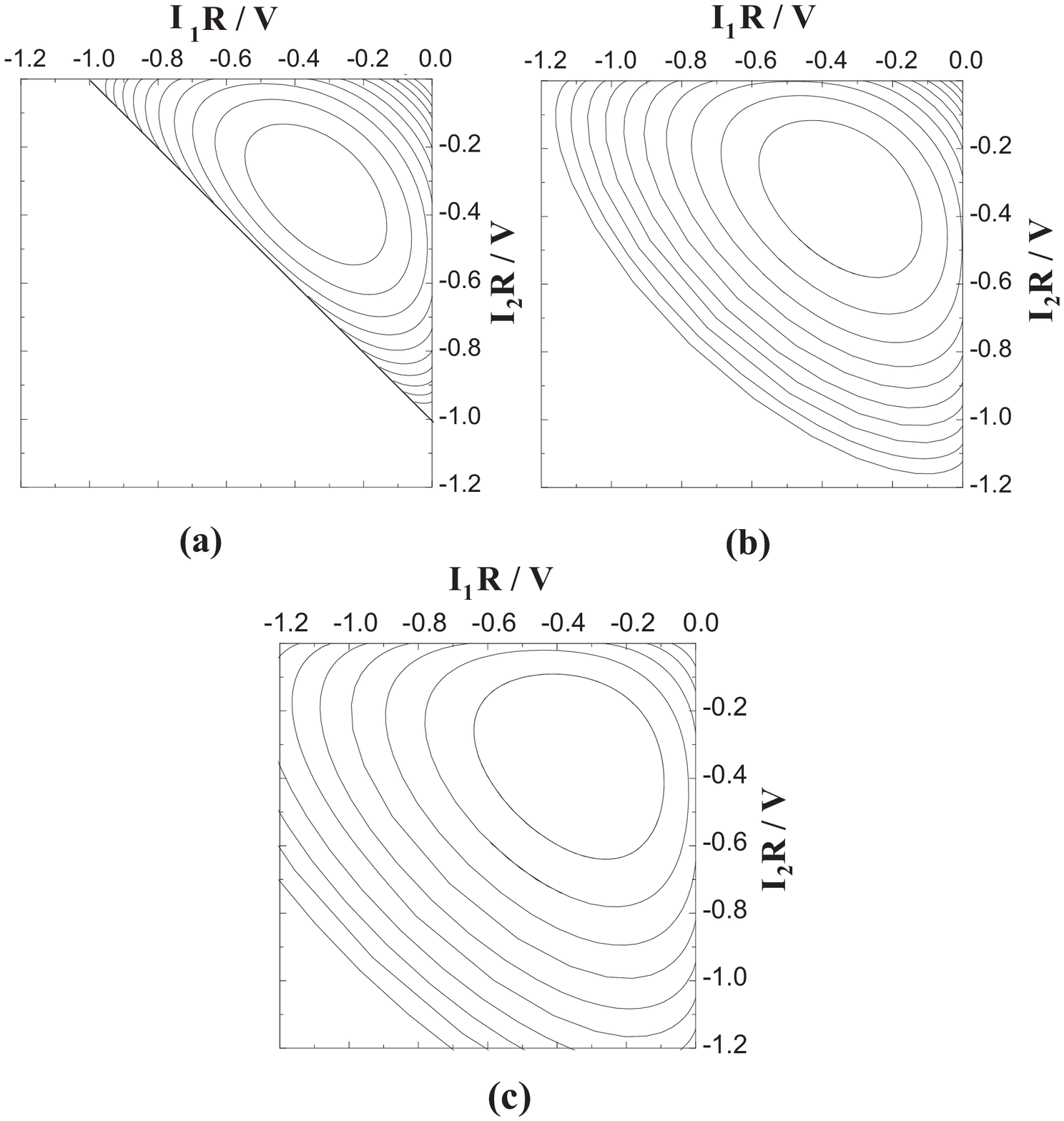}
\caption{The contour maps of the current distribution $\log[P(I_1,I_2)]$ in
the 3-terminal chaotic quantum dot for different configurations of connectors.
(a) - ballistic connectors, (b) - diffusive connectors, (c) - tunnel connectors.
\label{LnPCh} }
\end{center}
\end{figure*}

In the following we assume the shot noise regime $eV\gg kT$
when the thermal fluctuations can be disregarded. The energy
integration in~(\ref{FCS-Action}) becomes trivial, since $f_i(\epsilon)=0$ or 1,
and it is sufficient to consider only the case $V_1=V_2=0$, $V_3=V$. 
Any other possible setup can be reduced to the number of previous ones by appropriate 
subdividing a relevant energy strip. The results of these calculations are 
shown in Fig.~\ref{LogPCh} and \ref{LnPCh}. We see that the maximum of probability  
occurs at $I_1=I_2=-V/3R$, $I_3=2V/3R$. This simply reflects the usual Kirchoff rules. 
The current distribution $P(I_1,I_2)$ for a ballistic system is bounded. 
This stems from the fact, that each ballistic contact has a limited number
$N$ of open channels. Therefore any plausible current fluctuation in the
given terminal may not exceed the threshold value of a current $I_s = N V/R_Q$ via
the ballistic connector. Contrary to that, 
for the tunnel and diffusive type configurations any connector has an infinite number
of partially open transmission channels. Therefore the range of current fluctuations
is not bounded in this case. 

  From Fig.~\ref{LogPCh} and \ref{LnPCh}
it follows that the relative probabilities of big current fluctuations increase in 
the sequence ballistic$\rightarrow$diffusive$\rightarrow$tunnel. To reveal the
origin of this behavior we proceed by considering the shot-noise cross-correlations
$P_{ij}$, defined by Exp.~(\ref{Ps}). At zero frequency they can be found analytically
via the relation 
$P_{ij} = -\frac{e^2}{t_0} \frac{\partial^2}{\partial \chi_i \partial \chi_j} S\Big|_{\chi=0}$
Taking into account, that the action $S(\{\chi_i\})$, regarded as function of $\chi_i$,
is implicitly determined by Eq.~(\ref{FCS-Action}) via the set of equations~(\ref{mapping}),
we arrive to the following result
\begin{equation}
 P_{ij} = -G_{ij} (T+\Theta_4) + R^{-1}\alpha_i\alpha_j 
(\bar \Theta - \Theta_i - \Theta_j) + \delta_{ij} R^{-1} \alpha_i\Theta_i
\label{Pij}
\end{equation}
Here $\alpha_i = R/R_i$, $R^{-1} = \sum_i R^{-1}_i$,
\begin{equation}
  G_{ij} = R^{-1}(\alpha_i \alpha_j - \delta_{ij}\alpha_i)
\end{equation}
is a conductance matrix of the dot,
\begin{equation}
 \Theta_4 = \int d\varepsilon \bar f(1-\bar f), \quad
 \Theta_k = F_k \int d\varepsilon (f_k-\bar f)^2, \quad
 \bar\Theta = \sum_{i=1}^{3} \alpha_i \Theta_i 
\end{equation}
$F_k$ is a Fano factor of the $k$-th connector and 
$ \bar f(\varepsilon) = \sum_{i=1}^{3} \alpha_i f_i(\varepsilon)$ is the non-equlibrium
distribution function within the dot.

   In case of ballistic contacts $F_k = 0$ and the first two terms of the result~(\ref{Pij})
reproduce the expression for the noise power obtained with the use of "minimal correlation"
approach\cite{BlanterSukhor}. At zero temperature it also coincides with the result 
of random matrix theory~\cite{Langen}, as we expected from the our consideration
of scattering approach.
At equilibrium $P_{ij} = -2T G_{ij}$ in
accordance with the fluctuation-dissipation theorem. 
In the general situation each non-ideal connector with $F_k\neq 0$ gives the
additive contribution to the cross-correlation function, which is linear in $F_k$.
Thus we conclude that the current-current correlations in the chaotic quantum dot result
from two  contributions. The first one, proportional to $(T+\Theta_4)$,
corresponds to the cross-correlation function
$P_{ij}^{\rm b}$ of the chaotic cavity with ideal ballistic point contacts, which
stems from the fluctuation of the distribution function $\bar f$ within the dot.
The second contribution, proportional to $\Theta_i$, reflects the noise
due to connectors.
 
  To present the result for the cross-correlations it is also useful to introduce 
the ($3\times 3$) matrix $F$ with elements $F_{ij} = P_{ij}/eI_\Sigma$,
where $I_\Sigma = \sum_{i=1}^3 |I_i|$. The matrix $F$ is a generalization of the 
Fano factor for the multi-terminal system. It is symmetric and obeys the relation 
$\sum_{i=1}^3 F_{ik}=0$, which follows from the current conservation law.
At zero temperature for the symmetric setup, shown in Fig.~\ref{LogPCh}, it reads
as
\begin{equation} 
F = \frac{1}{36}\left(
\begin{array}{ccc}
4+3F   & -2      & -(2+3F) \\
-2       & 4+3F  & -(2+3F)  \\
-(2+3F)  & -(2+3F) &  4+6F
\end{array} \right)
\end{equation}
For a diffusive wire $F_D=1/3$ and for a tunnel junction $F_T=1$. Therefore
at fixed average currents through  connectors the Gaussian's currents 
fluctuations will increase in the sequence
ballistic$\rightarrow$diffusive$\rightarrow$tunnel. As it was mention before
similar behavior is also traced in the regime of the big  current 
fluctuations. The essential point here is that the cross-correlations 
always persist regardless the concrete construction of the connectors.

\end{subsection}
\end{section}

\begin{section}{FCS of Charge Transfer in Coulomb Blockade Systems}

   In the preceding sections we have considered the FCS of non-interacting
electrons. It was assumed that the conductance $G$ of the system is much greater
than the quantum conductance $G_Q=e^2/\hbar$. It is known, that under 
opposite condition, $G\leq G_Q$, the effects of Coulomb blockade 
become important. This motivates us to study the statistics of charge transfer
in the mesoscopic systems, placed in the limit of strong Coulomb interaction, 
$G\ll G_Q$.  The electrons dynamics in this Coulomb blockade limit
is fortunately relatively simple, since  
the evolution of the system is governed by a master equation.
The charge transfer is thus a classical stochastic
process rather than the quantum mechanical one. Nevertheless the FCS is by no means trivial.
In this section we elaborate the general approach to FCS in the systems, governed
by master equation.

  We begin this section by presenting the general model, which dynamics obeys the 
master equation. Further on we proceed with the proof
of the central result of this section for the FCS in the master equation approach. 
We illustrate the general scheme
by considering the statistics of charge transfer through the Coulomb blockade
island with 3 leads attached and compare the FCS in this case with the results
of the preceding section, i.e. for non-interacting electrons. 

\begin{subsection}{The general model}

The dynamics of various systems can  be described by master equation. 
For our purposes it is convenient to write it in the matrix form:
\begin{equation}
 \frac{\partial}{\partial t}|p\,(t)\rangle = -\hat{L} |p\,(t)\rangle
 \label{MasterEq}
\end{equation}
where each element $p_n(t)$ of the vector $|p(t)\rangle$ is the probability to
find the system in the  state $n$. The matrix elements of operator $\hat L$ are given by 
\begin{equation}
 L_{mn} = \delta_{nm}\gamma_n - \Gamma_{m\leftarrow n}, \quad
 \gamma_n = \sum_{m\ne n} \Gamma_{m\leftarrow n} \label{Lmatrix}
\end{equation}
Here $\Gamma_{n\leftarrow m}$  stands for the transition rate from the state $m$
to the state $n$,  $\gamma_n$ presents the total transition rate from the state $n$. 
The $\hat L$ operator thus defined always has a zero eigenvalue, the corresponding 
eigenvector being the stationary solution of the master equation. 

  Coulomb blockade mesoscopic systems always obey Eq.~({\ref{MasterEq}}).
The main advantage of the master equation approach
is a possibility of non-per\-tur\-ba\-tive treatment of the interaction effects. 
Below we first consider the master equation description of the general model
system. This will prepare us to the next section where we derive
the FCS method.

\begin{figure}[t]
\begin{center}
\includegraphics[scale=0.22]{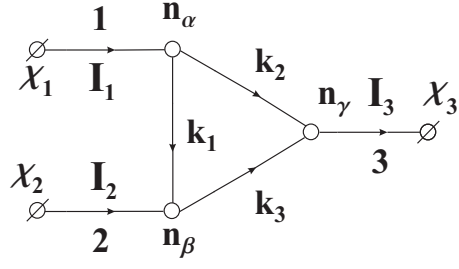}
\caption{The graph of the general model (See the main text). The terminals are connected
with the system via external junctions 1,2 and 3. The nodes $\alpha$, $\beta$ and $\gamma$
are either resonant levels or dots, linked with each other by internal junctions $k$'s.
The arrows denote the conventional direction of a current through each junction.
\label{MEGraph}
} 
\end{center}
\end{figure} 
 
   The possible physical realization of the general model includes
an array of Coulomb blockade quantum dots and a mesoscopic system with a number
of resonant levels. Like in the preceding sections, 
it is convenient schematically to present the system as a graph (see Fig.~\ref{MEGraph})
with each node $\alpha$ corresponding either to a single dot, a single resonant 
level or an external terminal. The line $k=(\alpha,\beta)$, connecting
the nodes $\alpha$ and $\beta$, is associated with a possible electron transfer.
Let $M$ be the total number of nodes in this graph and 
$L$ is a total number of lines.  For a many-dot systems each line $k$ corresponds 
to the tunnel junction. 
For systems with many resonant levels it corresponds to the
possible transition between different levels,
so that it does not necessary correspond to electron transfer in space. 
There are $N$ external 
junction $k=1\dots N$, ($N\le L$), connecting the terminals with the system.  The currents 
through these junctions are directly measurable and hence are of our 
interest.

  The macro- or microscopic state of the general model is given by a set of occupation
numbers $|n\rangle = |n_1, \dots, n_M\rangle$;  $n_\alpha$ is equal to any
integer for the array of quantum dots and refers to the excess charge on the
island $\alpha$; in case of many resonant level $n_\alpha$ denotes
the occupation number of a given level.
Owing to the fact that $\sum_n L_{nm} = 0$, the $\hat L$ operator has a zero
eigenvalue. There are the right, 
$|p_0\rangle$, and the left, $\langle q_0|$, eigenvectors corresponding to this
zero eigenvalue
\begin{equation}
 \hat L |p_0\rangle = 0, \qquad \langle q_0 |\hat L = 0
\label{Equilib}
\end{equation}
We assume that they are unique.
This means that the system does not get stuck in any metastable state.
The vector $|p_0\rangle$ gives the steady probability distribution and
$\langle q_0 | = (1, 1, \dots, 1)$. 

  It is also useful to present $\hat L$ operator in the form
\begin{equation}
\hat L = \hat \gamma - \hat \Gamma, \quad
\hat\Gamma = \sum_{k=1}^L (\hat\Gamma_k^{(+)} + \hat\Gamma_k^{(-)})
\end{equation}
where $\hat\gamma$ is the diagonal operator in the basis $|n\rangle$ of the system
configuration
and $\hat\Gamma_k^{(\pm)}$ is associated with the electron transfers through the
line $k=(\alpha,\beta)$:
\begin{equation}
 \hat\gamma = \sum_{\{n\}} |n\rangle\gamma(n) \langle n|, \quad
  \hat\Gamma_k^{(\pm)} = \sum_{\{n\}} |n'\rangle\Gamma_k^{(\pm)}(n) \langle n|
\label{GOperators}
\end{equation}
 The state $|n'\rangle = |n_1,\dots,n'_\alpha,\dots, n'_\beta,\dots, n_M \rangle$ results
from the state $|n\rangle$  by appropriate changing the corresponding occupation numbers:
$n'_\alpha = n_\alpha - \sigma_k$,  $n'_\beta = n_\beta + \sigma_k$, where $\sigma_k=\pm 1$
denotes the direction of the transition. 
   
\end{subsection}

\begin{subsection}{The FCS in the master equation}
In this section we derive the central result for the FCS of the charge transfer
in the system, which dynamics obeys the master equation. We will solve this
problem by making use of the property of the system, 
that its random evolution in time is the Markov  stochastic process.

  In what follows we will partially use notations of the book~\cite{vanKampen}. 
Let us consider the time interval $[-T/2, T/2]$. Suppose the system undergoes $s$
transitions at random time moments $\tau_i$, so that
\begin{equation}
 +T/2 > \tau_1 > \tau_2 > \dots > \tau_{s-1} > \tau_s > -T/2
\label{Times} 
\end{equation}
This gives an elementary random sample 
$\zeta_s=(\tau_1, k_1, \sigma_1; \dots ; \tau_s, k_s, \sigma_s)$. It corresponds to
the set of subsequent events, when at time $\tau_i$ the tunneling happens
via the junction $k_i$, $\sigma_i = \pm 1$ being the direction  of the transition.
The samples $\zeta_s$ constitute the set $\Omega$ of all possible random samples. 

 Then one defines the measure (or the probability) $d\mu(\zeta)$
at the set $\Omega$. For this purpose we may very generally introduce the
sequence of non-negative probabilities
$Q_s(\{\tau_i, k_i, \sigma_i\})\equiv  
Q(\tau_1,k_1,\sigma_1; \dots ;\tau_s, k_s, \sigma_s)\ge 0$
defined in $\Omega$ so that
\begin{equation}
 d\mu(\zeta) = Q_0 + 
 \sum_{s=1}^{+\infty} \sum_{\{k_i,\,\sigma_i\}}
Q_s(\{\tau_i, k_i, \sigma_i\})  d\tau_1 \dots d\tau_s 
\label{measure} 
\end{equation}
The functions $Q$ are normalized according to the condition
\begin{gather}
 \int_{\Omega}d\mu(\zeta) \equiv Q_0 + \nonumber \\  \sum_{s=1}^{+\infty}
\sum_{\{k_i,\,\sigma_i\}}\quad
\idotsint\limits_{T/2>\tau_1>\dots>\tau_s>-T/2}
Q_s(\{\tau_i, k_i, \sigma_i\}) \prod_{i=1}^s d\tau_i  = 1
\label{Norm}
\end{gather}
Each term in Exp.~(\ref{measure}) corresponds to the probability 
of an elementary sample $\zeta_s$.

  To accomplish the preliminaries, we remind the concept of 
a stochastic process. Mathematically speaking, it can be any integrable function
$\check A(t) \equiv A(t,\zeta)$ defined at the set
$\Omega$ and parametrically depending on time. It is sometimes convenient to omit the explicit 
$\zeta$ dependence. We will use a "check" in this case to stress that the quantity in 
question is a random variable.
Each stochastic process $A(t,\zeta)$ generates the sequence 
of time dependent functions
$\Bigl\{ A_0(t), \, A_1(t,\tau_1,k_1,\sigma_1), \dots $ 
$, A_s(t,\{\tau_i, k_i, \sigma_i\}) \Bigr\}$.
Its average ${\langle \check A(t) \rangle}_\Omega$ over the space $\Omega$ is defined as
\begin{gather}
{\langle \check A(t) \rangle}_\Omega = \int_{\Omega} A(t,\zeta) d\mu(\zeta) \equiv 
A_0(t) Q_0 + 
\sum_{s=1}^{+\infty}\sum_{\{k_i,\, \sigma_i\}} \label{Average} \\
\idotsint\limits_{T/2>\tau_1>\dots>\tau_s>-T/2}
A_s(t,\{\tau_i, k_i, \sigma_i\}) 
Q_s(\{\tau_i, k_i, \sigma_i\}) \prod_{i=1}^s d\tau_i  
\nonumber
\end{gather}
The analogous expression should be used, for instance, to define the correlations 
${\langle \check A(t_1) \check B(t_2) \rangle}_\Omega$
between any two stochastic processes.  

   For the subsequent analysis we define the random process ${\check I^{(k)}(t)}$, 
corresponding to the classical current through the external junction $k\le N$:
\begin{equation}
 I^{(k)}(t, \zeta_s) = \sum_{i=1}^{s} e\sigma_i\, \delta(t-\tau_i) \delta(k-k_i)
\label{Current}
\end{equation}
Here $\sigma_m$ is included to take into account the direction of the jump
and $\delta(k-k_i)\equiv \delta_{k,k_i}$ is the Kronecker $\delta$ symbol. Given this
definition at hand, we introduce the generating functional  $S[\{\chi_i(t)\}]$ 
depending on $N$ counting fields $\chi_i(\tau)$, each of them associated with 
a given terminal $i$:
\begin{equation}
\exp(-S[\{\chi_i(t)\}]) = {\left\langle \exp\Bigl\{ i \sum_{n=1}^N 
\int\limits_{-\infty}^{+\infty} d\tau \chi_n(\tau)\check I^{(n)}(\tau)/e
\Bigr\} \right\rangle}_\Omega
\label{Action}
\end{equation}
with the average defined by Eq.(\ref{Average}). 
Let us note the remarkable similarity of this classical expression with 
the quantum mechanical action~(\ref{QM_Action}).
As before, in the low-frequency
limit of current correlations one may use the time-independent counting fields $\chi_i$.
In this case the action $S[\{\chi_i\}]$ can be used to find the 
probability~(\ref{Def}) of $N_i$
electrons to be transferred through the terminal $i$ during the time interval $T$.

  The above definitions were rather general than constructive, since the
probabilities $Q$ have not been specified so far. To proceed, one has to relate
them to transition rates of the master equation. We assume that
at initial time $t=-T/2$ the system was in the state $\{n^{(s)}\}$. 
Then random sample $\zeta_s$ determines the evolution of 
charge configuration 
$\{n^{(s)}\}\rightarrow\{n^{(s-1)}\}\dots \{n^{(1)}\}\rightarrow\{n^{(0)}\}$
for subsequent moments of time.
The choice of $\zeta_s$ specifies that the transition between neighboring charge
states $\{n^{(i)}\}$ and $\{n^{(i-1)}\}$ occurs at time $\tau_i$ via the 
junction $k_i=(\alpha_i,\beta_i)$.
Therefore the sequence $\{n^{(i)}\}$ is given by the relation 
$n^{(i-1)}_{\alpha_i}=n^{(i)}_{\alpha_i}-\sigma_{k_i}$, 
$n^{(i-1)}_{\beta_i}=n^{(i)}_{\beta_i}+\sigma_{k_i}$,  and
$n^{(i-1)}_{\gamma}=n^{(i)}_{\gamma}$ for all $\gamma\ne\alpha_i$ and $\beta_i$.
To determine the probability $Q_s(\{\tau_i, k_i, \sigma_i\})$
we note that (i) the sample $\zeta_s$ constitutes the Markov chain 
(ii) the conditional probability of the system  to  remain at state
${n^{(i)}}$  between the times $\tau_{i+1}$ and  $\tau_i$ is proportional to 
$\exp[-\gamma(n^{(i)})(\tau_{i}-\tau_{i+1})]$; (iii) the probability that the transition
occurs via the junction $k_i$ during the time interval $d\tau_i$ at the moment $\tau_i$
is given by $\Gamma_{k_i}^{(\sigma_i)}(n^{(i)})d\tau_i$. 
These arguments suggest that $Q$'s have the form
\begin{gather}
 Q_0  = Z_0^{-1}\exp[-\gamma(n^{(s)})T]\label{Qprob} \\
 Q_s(\{\tau_i, k_i, \sigma_i\})   =  Z_0^{-1}
 \exp[-\gamma(n^{(0)})(T/2-\tau_1)]\Gamma_{k_1}^{(\sigma_1)}(n^{(1)}) \nonumber \\
 \exp[-\gamma(n^{(1)})(\tau_1-\tau_2)]\Gamma_{k_2}^{(\sigma_2)}(n^{(2)})\dots 
 \exp[-\gamma(n^{(s-1)})\nonumber \\ (\tau_{s-1}-\tau_{s})] 
   \Gamma_{k_s}^{(\sigma_s)}(n^{(s)})
 \exp[-\gamma(n^{(s)})(\tau_{s}+T/2)] \nonumber
\end{gather}
where the constant $Z_0$ should be found from the normalization condition~(\ref{Norm}). 
As we will see below, $Z_0=1$.

 The above correspondence between the random Markov chain $\zeta_s$ and the probabilities 
$Q$'s~(\ref{Qprob}) allows one to evaluate the generating function~(\ref{Action}). 
By definition~(\ref{Current}) for any given $\zeta_s$ we have
\begin{gather}
\exp\Biggl\{i\sum_{n=1}^N 
\int\limits_{-\infty}^{+\infty} d\tau \chi_n(\tau) I^{(n)}(\tau, \zeta_s)/e\Biggr\} = 
\prod_{i=1}^s \exp\{i\sigma_i\,\chi_{k_i}(\tau_i)\} \nonumber
\end{gather}
It is assumed here that $\chi_{k_i}=0$ if the transition occurs via internal junction, 
$k_i>N$, thus no physically measurable current is generated in this case.
The averaging of the latter expression
over all possible configurations $\Omega$ with the weight $d\mu(\zeta)$
yields 
\begin{gather}
 Z[\{\chi_i(\tau)\}] \equiv\exp(-S[\{\chi_i(\tau)\}]) = Q_0 + \label{Zchi}\\
\sum_{s=1}^{+\infty}
\sum_{\{k_i,\,\sigma_i\}}\quad
\idotsint\limits_{T/2>\tau_1>\dots>\tau_s>-T/2}
Q_s^{\chi}(\{\tau_i, k_i, \sigma_i\}) \prod_{i=1}^s d\tau_i \nonumber
\end{gather}
The resulting expression resembles the normalization condition (\ref{Norm}).
Here the $\chi$-dependent functions $Q_s^{\chi}(\{\tau_i, k_i, \sigma_i\})$ are
defined similar to probabilities~(\ref{Qprob}) with the only crucial difference 
that the rates 
$\Gamma_{k}^{(\sigma)}(n)$ should be replaced by 
$\Gamma_{k}^{(\sigma)}(n)\exp\{i\sigma_k\,\chi_k(\tau_k)\}$ if $k\le N$.

The expression~(\ref{Zchi}) can be written in the more compact and elegant way.
For that, we introduce the $\chi$-dependent linear operator $\hat L_{\chi}$ defined as
\begin{eqnarray}
\hat L_{\chi}(\tau) &=& \hat \gamma - \hat \Gamma_{\chi}(\tau), 
\label{Lchi} \\
\hat\Gamma_{\chi}(\tau) &=& \sum_{k=1}^N 
(\hat\Gamma_k^{(+)}e^{i\chi_k(\tau)} + \hat\Gamma_k^{(-)}e^{-i\chi_k(\tau)}) 
+\sum_{k=N+1}^L (\hat\Gamma_k^{(+)} + \hat\Gamma_k^{(-)})
\nonumber
\end{eqnarray}
In line with consideration above we multiplied each operator 
$\hat\Gamma_k^{(\pm)}$ ($k=1\dots N$), that 
corresponds to the transition through the external junction, by 
an extra $\chi$-dependent factor $e^{i\chi_k(\tau)}$.  
The diagonal part and internal transition operators $\hat\Gamma_k^{(\pm)}$ with $k>N$
remained unchanged. Then we consider the evolution operator
$\hat U_{\chi}(t_1, t_2)$ associated with (\ref{Lchi}). Since  
$\hat L_{\chi}(\tau)$ is in general time-dependent, $\hat U_{\chi}(t_1, t_2)$  is given 
by the time-ordered exponent
\begin{equation}
 \hat U_{\chi}(t_1, t_2) = T_\tau \exp\Bigl\{-\int_{t_2}^{t_1}
 \bigl(\hat\gamma(\tau)- \hat\Gamma_{\chi}(\tau)\bigr)d\tau \Bigr\}
\label{TimeExp}
\end{equation}
The similar construction is widely used in quantum statistics. 
The difference in the present case 
is that the operator $\hat U_{\chi}(t_1, t_2)$  at $\chi=0$  gives
the evolution of probability rather than the amplitude of probability.
 
  With the use of evolution operator (\ref{TimeExp}) the generating function 
(\ref{Zchi}) can be cast into the form
\begin{equation}
Z[\{\chi_i(\tau)\}]  = \langle q_0|\hat U_{\chi}(T/2, -T/2)|n_s\rangle 
\label{UAction}
\end{equation}
To prove it we argue as follows. We exploite the fact 
that $\hat\gamma(\tau)$ and $\hat\Gamma(\tau)$ commute under the sign of 
time-ordering 
in Eq. (\ref{TimeExp}) and regard $\hat\Gamma(\tau)$ as a perturbation. 
This gives the matrix element $\langle q_0|\hat U_{\chi}(T/2, -T/2)|n_s\rangle$ 
in the form of series
\begin{gather}
\langle q_0|\hat U_{\chi}(T/2, -T/2)|n_s\rangle  = \langle q_0| e^{-\hat\gamma T}|p_0\rangle +
\label{USeries} \\
\sum_{s=1}^{+\infty}
\langle q_0|T_\tau \exp\Bigl\{-\int_{-T/2}^{T/2} \hat\gamma(\tau) d\tau \Bigr\} 
\sum_{{\bf k}_s {\bm\sigma}_s} \quad \idotsint\limits_{T/2>\tau_1>\dots>\tau_s>-T/2} \nonumber 
\\
\hat\Gamma_{k_1}^{(\sigma_1)}(\tau_1)e^{i\sigma_1\chi_{k_1}(\tau_{k_1})}
 \dots \hat\Gamma_{k_s}^{(\sigma_s)}(\tau_s)e^{i\sigma_s\chi_{k_s}(\tau_{k_s})}|p_0\rangle 
\prod_{i=1}^s d\tau_i
\nonumber
\end{gather}
It follows from the definition (\ref{Qprob}) that each term in this series
corresponds to the function $Q_s^{\chi}(\{\tau_i, k_i, \sigma_i\})$, namely
\begin{gather}
Q_0  =  \langle q_0|e^{-\gamma T}|n_s\rangle \nonumber \\
Q_s^{\chi}(\{\tau_i, k_i, \sigma_i\}) = 
\langle q_0|T_\tau \exp\Bigl\{-\int_{-T/2}^{T/2} \hat\gamma(\tau) d\tau \Bigr\} \\
\hat\Gamma_{k_1}^{(\sigma_1)}(\tau_1)e^{i\sigma_1\chi_{k_1}(\tau_{k_1})}
 \dots \hat\Gamma_{k_s}^{(\sigma_s)}(\tau_s)e^{i\sigma_s\chi_{k_s}(\tau_{k_s})}
 \nonumber
|n_s\rangle 
\end{gather}
Therefore Exp.~(\ref{USeries}) and  (\ref{UAction}) are reduced to the previous 
result (\ref{Zchi}). This completes the proof.  Note, that
owing to the property (\ref{Equilib}),
$Z_0 = \langle q_0|\exp(-T\hat L)|n_s\rangle=1$  identically 
at $\chi=0$. Therefore the probabilities (\ref{Qprob}) are correctly normalized.

  The Exp.~(\ref{UAction}) for the generating function $Z[\{\chi_i(t)\}]$ 
depends on the initial state  $|n_s\rangle$ of the system. 
It can be shown that the choice
of $|n_s\rangle$ does not affect the final results. We assume 
that $\chi_k(t)\to 0$ when $t\to -T/2$. Physically, it means that the 
measurement is limited in time. To be specific one may assume that 
$\chi_k(t)=0$ when $-T/2 < t < -T/2+\Delta t$ and $\chi_k(t)\neq 0$ if $t>-T/2+\Delta t$.
If the time interval $\Delta t$
is sufficiently large as compared with the typical transition time $\Gamma^{-1}$,
then the system will reach the steady state during this period of time. The latter follows
from the fact that $\exp(-\hat L\,\Delta t )|n_s\rangle \to |p_0\rangle$ 
when $\Delta t \gg \Gamma^{-1}$. Thus one can substitute 
$|n_s\rangle$ to $|p_0\rangle$ in Exp.~(\ref{UAction}). Assuming also the 
limit $T\to \infty$, we arrive to the main result of this section
\begin{equation}
 \exp(-S[\{\chi_i(t)\}]) = \langle q_0| T_\tau \exp\Bigl\{-\int\limits_{-\infty}^{+\infty}
 \hat L_{\chi}(\tau) d\tau \Bigr\}|p_0\rangle
\label{TimeS}
\end{equation}
We see that the generating function can be written in the form of the averaged evolution 
operator. This operator corresponds to master equation with the rates
modified by the counting fields $\chi_i(\tau)$.

   Further simplification is valid in the low frequency limit of the current 
correlations, $\omega \ll \Gamma$.
(Here $\Gamma$ is a typical transition rate in the system.)
This allows to set $\chi_k(t) = \chi_k$ when $0\le t\le t_0$ and $\chi_k(t) = 0$
otherwise, where $t_0$ is a time of measurement.
The action (\ref{TimeS}) then reduces to the 
\begin{equation}
S(\{\chi_i\}) = t_0 \Lambda_{\rm min}(\{\chi_i\}) 
\label{Lambda}
\end{equation}
where $\Lambda_{\rm min}(\{\chi_i\})$ is the minimum eigenvalue of the operator $\hat L_\chi$.
Thus the problem of statistics in the Coulomb blockade regime, provided the transition rates in 
the system are known, is merely a problem of the linear algebra.

\end{subsection}

\begin{subsection}{The statistics of charge transfer in Coulomb blockade island.}
  
\begin{figure}[t]
\begin{center}
\includegraphics[scale=0.3]{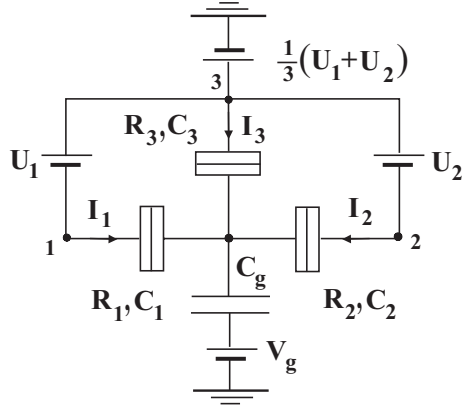}
\caption{The equivalent circuit of the three-terminal Coulomb blockade island.
The voltages $U_{1(2)}$ 
are used to control the bias between the 3d and the 1st (the 2nd) terminals:
$U_{1(2)}=V_3-V_{1(2)}$. The third terminal is biased at voltage $V_3=(U_1+U_2)/3$ with
respect to the ground. This setup assures the condition
$V_1+V_2+V_3 = 0$. Then gate voltage, $V_g$, can be used to control the
offset charge $q_0=C_g V_g$ on the island. 
\label{CBDot}
}
\end{center}
\end{figure}

In this subsection we use the developed method to study the
current statistics in the three-terminal Coulomb blockade island. Its equivalent circuit
is shown in Fig.~\ref{CBDot}. This circuit is an
extension of the usual set-up of a conventional single electron transistor~\cite{IngoldNaz}.  

   The island is in the Coulomb blockade regime, $R_k \gg R_Q = 2\pi\hbar/{e^2}$.
In order to observe the Coulomb blockade effect the condition 
$k_B T \ll E_c = e^2/2C_{\Sigma}$ should be also satisfied.
Here $E_c$ is the charging
energy of the island and $C_{\Sigma} = \sum_{i=1}^3 C_k + C_g$.
We assume the temperature to be rather high, 
$k_B T \gg \Delta E$, with $\Delta E$ being the mean level spacing in the island, 
so that the discreteness of the energy spectrum in the island is not important.
We also disregard the possible effects of co-tunneling.

Under the above conditions the 3-terminal island is  described
by the "orthodox" Coulomb blockade theory. 
In this theory the macroscopic state of Coulomb blockade island 
is uniquely determined 
by the excess charge $Q=ne$, which is quantized in terms of electron 
charge $(-e)$.  The charge $Q$ can be changed only by $\pm e$ in course of one tunneling event. 
Therefore the master equation connects the given macroscopic
state $n$ with the neighboring  states $n \pm 1$ only.  The corresponding rates 
$\Gamma_{n\pm 1\leftarrow n}$ of these transitions are equal to the sum
of tunneling rates across all junctions: 
$\Gamma_{n\pm 1\leftarrow n } = \sum_{k=1}^N \Gamma_{n\pm 1\leftarrow n }^{(k)}$.
The  tunneling rate $\Gamma_{n\pm 1\leftarrow n }^{(k)}$ across the junction $k$ 
can be expressed via the electrostatic energy difference $\Delta E_{n\pm 1\leftarrow n }^{(k)}$
between the initial ($n$) and
final $(n\pm 1)$ configurations
\begin{equation}
 \Gamma_{n\pm 1\leftarrow n }^{(k)} = \frac{1}{e^2 R_k}
\frac{ \Delta E_{n\pm 1\leftarrow n }^{(k)} }
{1-\exp[-\Delta E_{n\pm 1\leftarrow n }^{(k)}/k_B T]}
\label{QRates}
\end{equation}
The evaluation of  $\Delta E_{n\pm 1\leftarrow n }^{(k)}$ can be done along the same lines
as in the case of single electron transistor~\cite{IngoldNaz}.
   
 According to the general result~(\ref{Lambda}) of the preceding subsection,  
one can find the FCS of the charge transfer through the island, 
by evaluating the minimum eigenvalue
$\Lambda_{\rm min}$ of the matrix $\hat L_\chi$. In case under consideration
this problem is reduced to the eigenvalue problem of the three-diagonal matrix:
\begin{equation}
 (\Lambda - \gamma_n)p_n + \Gamma_{n \leftarrow n+1 }^\chi p_{n+1} +
 \Gamma_{n \leftarrow n-1 }^\chi p_{n-1} = 0
\label{System}
\end{equation}
where $\gamma_n = \Gamma_{n \leftarrow n-1 } + \Gamma_{n \leftarrow n+1 }$, and 
$\Gamma_{n \leftarrow n\pm 1}^\chi = \sum_{k=1}^N \Gamma_{n\pm 1\leftarrow n}^{(k)} e^{\pm 
i\chi_k}$.
The index $+(-)$ corresponds to electron transition from (to) the island.

  To assess the FCS we have treated the related linear problem~(\ref{System}) numerically. 
We restrict the consideration to sufficiently 
low temperatures $k_B T\ll E_c$, 
so that the temperature dependence in rates~(\ref{QRates}) is non-essential. In this case
$\Gamma_k^{(\pm)}(n) = \Delta E_{n\pm 1\leftarrow n}^{(k)}/(e^2 R_k)$ when
$\Delta E_{n\pm 1\leftarrow n}^{(k)}\geq 0$ and $\Gamma_k^{(\pm)}(n) = 0$ 
otherwise. We have also assumed that $U_2>U_1$ (See Fig.~\ref{CBDot}).
The corresponding $\chi$-dependent rates can be found from 
Exp.~(\ref{QRates}) and (\ref{Lchi}) and read as follows
\begin{gather}
 \Gamma_{n+1\leftarrow n}^{\chi} = \Gamma_{3}^{(+)}(n)e^{i\chi_3} + 
 \Gamma_{1}^{(+)}(n)\,\theta(q-1/2-n)e^{i\chi_1} \nonumber \\
 \Gamma_{n-1\leftarrow n}^{\chi} = \Gamma_{2}^{(-)}(n)e^{-i\chi_2} + 
 \Gamma_{1}^{(-)}(n)\,\theta(n-q-1/2)e^{-i\chi_1} 
\label{rates3D}
\end{gather}
where
\begin{equation}
\Gamma_{k}^{(\pm)}(n) = a_k^{(\pm)} \mp \left(n+ \frac{C_g V_g}{e} \pm  \frac{1}{2}\right)
\frac{1}{R_k\, C_\Sigma}  
\end{equation}
and
\begin{gather}
a_3^{(+)} = \frac{\tilde C_1 U_1 + \tilde C_2 U_2}{e R_3\, C_\Sigma}, \quad 
a_2^{(-)} = \frac{ (\tilde C_1+ \tilde C_3) U_2-\tilde C_1 U_1}{e R_2\, C_\Sigma} \nonumber \\
a_1^{(\pm)} = \pm \frac{ \tilde C_2 U_2 - (\tilde C_3+ \tilde C_2) U_1}{e R_1\, C_\Sigma} \nonumber
\end{gather}
The effective capacitances $C_k$ are defined as $\tilde C_k = C_k + C_g/3$
and the point $q$ is given by the relation
\begin{equation}
  e q(U_1,U_2,V_g) = 
   \tilde C_2 U_2 - (\tilde C_3+ \tilde C_2) U_1 -C_g V_g  
\end{equation}
The value $q$ is non-integer in general. It satisfies the condition
$\Gamma_1^{(-)}(q+1/2)=\Gamma_1^{(+)}(q-1/2)=0$. 
The dimension of the $\hat L_\chi$-matrix is equal to $n_{\rm max}-n_{\rm min}$,
where $n_{\rm max}$($n_{\rm min}$) can be found from the conditions $\Gamma_3^{(-)}(n)\ge 0$
($\Gamma_1^{(+)}(n)\ge 0$).
The value $e\,n_{\rm max}$, ($e\,n_{\rm min}$) gives the maximum (minimum) charge 
that can be in the island for a given voltages $U_1$, $U_2$ and $V_g$.

  We can see from the Exp.~(\ref{rates3D}) that there are four elementary processes of charge
transfer in the system at low temperatures, each of them being associated with the pre-factor
$e^{\pm i\chi_k}$. The factors $e^{i\chi_3}$ and $e^{-i\chi_2}$  correspond to
the charge transfer from the third terminal into the island and from the island into the second 
terminal, respectively. Hence, the random current through the 3d (2nd) junctions always has the 
positive (negative) sign.  Two factors $e^{\pm\chi_1}$ stem from the charge transfer
through the first junction in the direction either from the island into the first contact or
vice versa. Therefore the current $I_1$  fluctuates in both directions.

\begin{figure}[t]
\begin{center}
\includegraphics[scale=0.3]{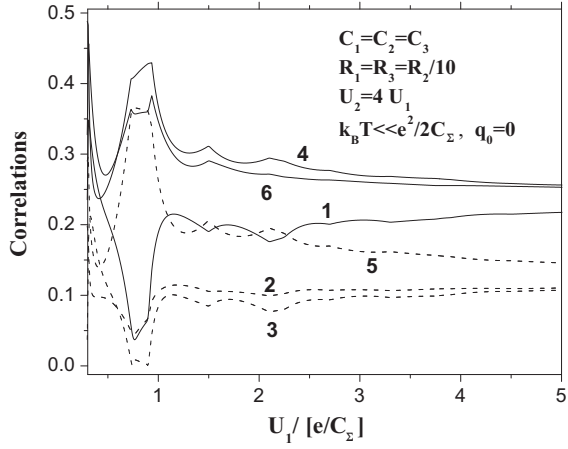}
\caption{ The matrix $F$ of auto- and cross- shot noise correlation versus voltage $U_1$
for the 3-terminal quantum dot setup.
Parameters are shown on the plot.
(1) - $F_{11}$, (2) - $|F_{12}|$, (3) - $|F_{13}|$, 
(4) - $F_{22}$, (5) - $|F_{23}|$, (6) - $F_{33}$.  }
\label{NoiseU}
\end{center}
\end{figure}

\begin{figure}[t]
\begin{center}
\includegraphics[scale=0.275]{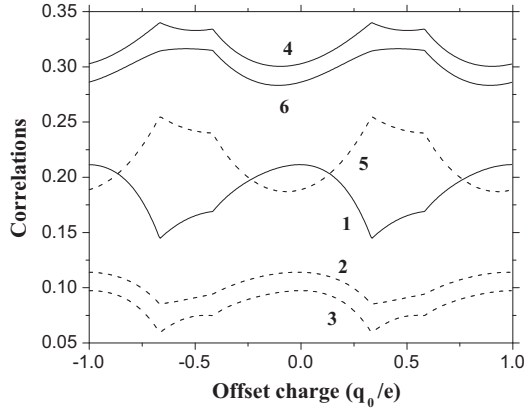}
\caption{ The matrix $F$ of auto- and cross- shot noise correlation versus the offset charge
for the 3-terminal quantum dot setup. Parameters are the same as on the Fig.~\ref{NoiseU}. 
The voltage
$U_1 = U_2/4 = 1.25 e/C_{\Sigma}$.
(1) - $F_{11}$, (2) - $|F_{12}|$, (3) - $|F_{13}|$, 
(4) - $F_{22}$, (5) - $|F_{23}|$, (6) - $F_{33}$.  
\label{NoiseQ}}
\end{center}
\end{figure}

  Let us consider the shot noise correlations in the system. 
In Fig.~\ref{NoiseU} we give the illustrative example of the voltage dependence of 
the shot noise correlations $F_{km}$ for a certain choice of parameters. 
The definition of matrix $F_{km}$ is the same as we have used in the end of section 3.
The Coulomb blockade features are strongly pronounced for an asymmetric setup only.
The results shown in Fig.~\ref{NoiseU} correspond
to $R_1=R_3=R_2/10$, $\tilde C_1 = \tilde C_2 =\tilde C_3$
and $U_2/U_1=4$. In Fig.~\ref{NoiseQ} we show the dependence of the
shot noise correlations on the offset charge for the same set of parameters and 
the value of $U_1=1.25\,e/C_\Sigma$. The special points of both these 
dependences occur when either $n_{\rm min}$, $n_{\rm max}$ or the integer part
of $q$ are changed by $\pm 1$. As the result one observes multi-periodic Coulomb blockade
oscillations in the offset charge dependences.

\begin{figure}[t]
\begin{center}
\includegraphics[scale=0.35]{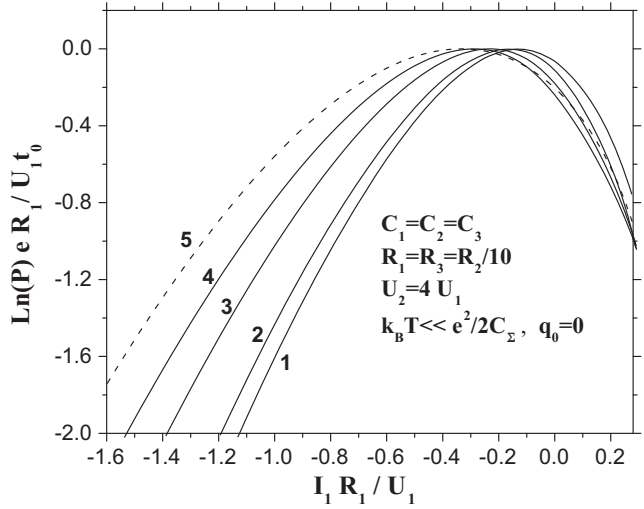}
\caption{ The logarithm of current distribution $\ln P(I_1, I_2)$ in the 3-terminal quantum dot 
as a function of current  $I_1$, under condition $I_2=\langle I_2 \rangle$. 
Parameters are shown on the plot. (1) - $U_1=1.25\,e/C_\Sigma$, (2) - $U_1=2.0\,e/C_\Sigma$, 
(3) - $U_1=4.0\,e/C_\Sigma$, (4) - $U_1=10.0\,e/C_\Sigma$; curve (5) corresponds to the
non-interacting regime. 
\label{LogCBL}}
\end{center}
\end{figure}

  We now proceed with the evaluation of the FCS. The action $S(\{\chi_i\})$
has been calculated with the use of (\ref{Lambda}).
To find the probability distribution $P$ we have evaluated the 
Fourier transform (\ref{Def}) in the saddle point approximation. 
It is applicable here, since we consider the low frequency limit only, $\omega\ll \Gamma$.
In this limit both action $S$ and number of transmitted particles $N_i=I_i t_0/e \gg 1$. 
Due to the current conservation
$\sum_k I_k$ = 0, only two currents are independent and
the action $S(\{\chi_i\})$ depends on the differences
$\chi_{ij} = \chi_i-\chi_j$ only. In what follows we have chosen $I_1$ and $I_2$ as
the independent variables to plot the logarithm of probability $\ln P(I_1,I_2)$. 
In the saddle point approximation, 
with the exponential accuracy, it is given by $P(I_1,I_2) \sim e^{-\Omega(\chi^*)}$.
Here $\chi^*$ is a saddle point of the function  
$\Omega(\chi) = S(\chi)+ i\chi_1 I_1 t_0/e + i\chi_2 I_2 t_0/e $. The
results for $\ln P(I_1,I_2)$ are shown in Figs.~\ref{LogCBL} and \ref{ContourB}(a). 
From the contour map
on Fig.~\ref{ContourB}(a) we see that $P(I_1,I_2)$ is non-zero in the region $I_1<0$, $I_2<0$
and in the region $I_1\le |I_2|$ provided $I_1>0$ and $I_2<0$. 
This range of plausible
current fluctuations stems from the $\chi$-dependence of rates (\ref{rates3D}).
Any current fluctuation automatically satisfies the constrain $\sum_k I_k$ = 0
and conditions $I_2<0$ and $I_3>0$. 

\begin{figure}
\begin{center}
\includegraphics[scale=0.4]{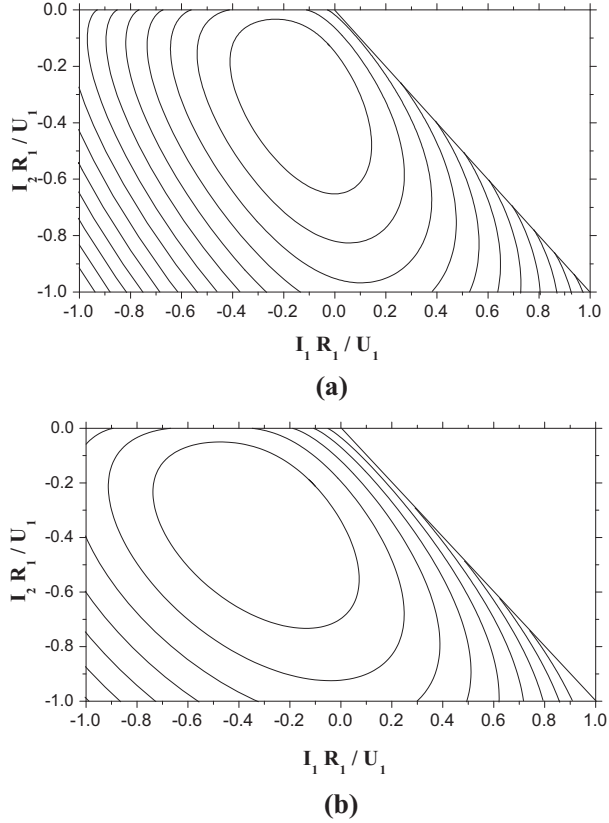}
\caption{The contour maps of the current distribution $\log[P(I_1,I_2)]$ in
the 3-terminal quantum dot. $U_1=U_2/4.0 = 1.25\,e/C_\Sigma$. 
(a) - Coulomb blockade island.  Parameters are 
the same as in Fig.~\ref{LogCBL}. (b) - Chaotic quantum dot. 
\label{ContourB}}
\end{center}
\end{figure}


   Before discussing the results, let us set the reference point for
such discussion. This reference will be the results of the previous section.
We consider the FCS in the three-terminal chaotic quantum dot when its 
contacts are tunnel junctions with resistances 
$R_k^{-1} \gg {e^2}/\pi\hbar$. In this limit the effects of interaction are 
negligible and electrons are scattered independently at different energies. 
Provided $U_2>U_1$, the generating function $S(\{\chi_i\})$ in the given case is a
sum of the two independent processes~(\ref{FCS-Action})
\begin{equation}
 S(\chi_1,\chi_2,\chi_3) = S_1(\chi_1,\chi_2,\chi_3) + S_2(\chi_1,\chi_2,\chi_3) 
\label{SNoninter}
\end{equation}
Here 
\begin{gather}
 S_1(\chi_1,\chi_2,\chi_3) = \frac{U_1 t_0}{2e}\Bigl\{ G_1 + G_2 + G_3  - \nonumber \\ 
\sqrt{ (G_1 + G_2 - G_3 )^2 + 
4 G_3 e^{i\chi_3} ( G_1 e^{-i\chi_1} + G_2 e^{-i\chi_2} ) } \Bigr\} \nonumber
\end{gather}
\begin{gather}
 S_2(\chi_1,\chi_2,\chi_3) = \frac{(U_2 - U_1) t_0}{2e}\Bigl\{ G_1 + G_2 + G_3  - \nonumber \\ 
\sqrt{ (G_1 + G_3 - G_2 )^2 + 
4 G_2 e^{-i\chi_2} ( G_1 e^{i\chi_1} + G_3 e^{i\chi_3} ) } \Bigr\} \nonumber 
\end{gather}
and $G_k = R_k^{-1}$ are the conductances  of the junctions.

The logarithm of probability $\ln P_0(I_1, I_2)$, evaluated with the use of 
statistics~(\ref{SNoninter}), is shown by the dashed line in Fig.~\ref{LogCBL}.
Its contour map for the same values of parameters is separately presented
in Fig.~\ref{ContourB}(b).  The maximum of $\ln P_0(I_1, I_2)$, as expected, occur at 
$\bar I_1 = \bar I_2 = U_1/3 R_1$.  

Comparing the FCS in the Coulomb blockade and non-interacting limits
we can draw the following conclusions.
In spite of the different re\-gimes, we see that 
the qualitative dependence of probabilities versus the currents is similar for both
cases. The probability distribution in both cases has a single maximum,
corresponding to the average values of currents. The tails of distribution
are essentially non-Gaussian both in the weak and strong interacting limit.
The statistics approaches to the Gaussian-type one in the strong Coulomb
blockade limit only, when the applied voltage to the system is only few above
the Coulomb blockade threshold. (See curves 1 and 2 in Fig.~\ref{LogCBL})
At higher applied voltages the probability distribution has a tendency to 
approach to the current distribution of the non-interacting system.
However, they never become identical, even in the limit $U_{1,2}\gg e/C_\Sigma$.
(Curves 4 and 5). The same is true for the shot-noise correlations.
Generally, we conclude that the Coulomb interaction always suppresses the relative 
probabilities of big current fluctuations. 
This stems from the fact that any big current fluctuation 
in Coulomb blockade dot is related with the
large accumulation (or depletion) of the charge on the island. The latter process
results in the excess of electrostatic energy. Therefore
the relative probability of such fluctuation is decreased, as compared to
the probability of the similar current fluctuation in the non-interacting regime.

\end{subsection}

\end{section}

\begin{section}{The equivalence of scattering and master equation approaches to the FCS}

\begin{figure}[t]
\begin{center}
\includegraphics[scale=0.25]{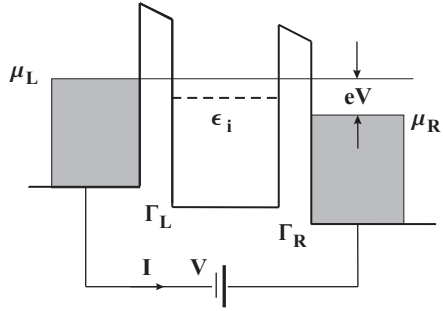}
\caption{ 
The single resonant level system, formed by the two tunnel barriers. The resonant
level in the quantum well is shown by the dashed line.
\label{Level}}
\end{center}
\end{figure}

In this section we evaluate the FCS in the generic case of a single resonant
level, shown in Fig.~\ref{Level}.  We consider only the non-interacting spinless electrons
and demonstrate the equivalence of scattering and master equation approaches 
to the FCS in the framework of this model.

   We start by considering the FCS in the master equation framework. 
It is applicable, provided the applied voltage or the temperature are not
too low, i.e. ${\rm max}\{eV,k_B T\}\gg \hbar \Gamma_{L(R)}$. Here $\Gamma_{L(R)}$
denote the quantum-mechanical tunneling rates from the left (right) electrode onto
the resonant level. The system can be found in the two microscopic states only:
one with no electron on the level, and another with a single electron. 
Then the transition rates involved reads as
 \begin{eqnarray}
 \Gamma_{1\leftarrow 0} &=& \Gamma_L f_L(\epsilon_i) + \Gamma_R f_R(\epsilon_i) 
\label{Rates_non_int} \\
 \Gamma_{0\leftarrow 1} &=& \Gamma_L [1-f_L(\epsilon_i)] + \Gamma_R [1-f_R(\epsilon_i)] 
\nonumber
\end{eqnarray}
Here the indices $\{0\}$ and $\{1\}$ denote the microscopic state with no and one
electron on the level.
Fermi function $f_{L(R)}(\epsilon) = (1+\exp[(\epsilon-\mu_{L(R)})/kT])^{-1}$
accounts for the filling factor in the left (right) lead and $\epsilon_i$ is the position of 
the resonant level. 

Following the definition~(\ref{Lchi}) and the expression for the rates~(\ref{Rates_non_int}),
the $\hat L_\chi$-matrix of the single resonant level model 
reads as 
\begin{equation}
\label{L_non_int}
\hat L_{\chi} = \left( 
\begin{array}{cc}
 \Gamma_{1\leftarrow 0}  & -  \Gamma_{0\leftarrow 1}(\chi) \\
 -\Gamma_{1\leftarrow 0}(\chi)  & \Gamma_{0\leftarrow 1} 
\end{array} \right)
\end{equation}
where
\begin{gather}
\label{Rates_chi}
\Gamma_{1\leftarrow 0}(\chi) = \Gamma_L f_L e^{-i\chi_1} + \Gamma_R f_R e^{-i\chi_2}  \\
\Gamma_{0\leftarrow 1}(\chi) = \Gamma_L (1-f_L)e^{i\chi_1} + \Gamma_R (1- f_R)e^{i\chi_2}
\nonumber
\end{gather}
  Evaluating the minimum eigenvalue of this matrix we obtain the current statistics
in the following form
\begin{gather}
 S(\chi)  = \frac{t_0}{2}\Bigl\{ \Gamma_L + \Gamma_R -\sqrt{{\cal D}(\chi)}  \Bigr\}\label{SME} \\
{\cal D}(\chi) = (\Gamma_L + \Gamma_R)^2 +  4\Gamma_L\Gamma_R 
\left[ f_{(-)}(\epsilon_i)( e^{-i\chi} -1) +  f_{(+)}(\epsilon_i)( e^{i\chi} -1) \right] 
\nonumber
\end{gather}
Here $f_{(-)}(\epsilon_i) =f_L(\epsilon_i)[1-f_R(\epsilon_i)]$, 
$f_{(+)}(\epsilon_i) =f_R(\epsilon_i)[1-f_L(\epsilon_i)]$ and $\chi = \chi_1-\chi_2$.

  Since the electrons above are assumed to be non-interacting, one might
have come to the same result in the framework of the pioneering approach by
Levitov {\it et.al.}\cite{LLL}. We will show now that it is indeed the case.

 According to Ref.~\cite{LLL} the general expression for the current statistics through a single
contact reads as 
\begin{gather}
 S(\chi) = -\frac{t_0}{2\pi}\sum_n \int d\epsilon\, {\rm ln}\Bigl\{1+ T_n(\epsilon)\times
\label{SLLL} \\
\Bigl( f_L(\epsilon)[1-f_R(\epsilon)]( e^{-i\chi} -1) + 
f_R(\epsilon)[1-f_L(\epsilon)]( e^{i\chi} -1) \Bigr)  
\Bigr\} \nonumber
\end{gather}
It is valid for any two-terminal geometry provided the region between two electrodes
can be described by the one-particle scattering approach.
$T_n(\epsilon)$ is a set of transmission eigenvalues which are in
general energy-dependent. For a resonant level there is a single
resonant transmission eigen-value $T_r(\epsilon)$, its energy
dependence being given by the Breit-Wigner formula
\begin{equation}
 T_r(\epsilon) = \frac{\Gamma_L \Gamma_R}{ (\epsilon - \epsilon_i)^2 + 
(\Gamma_L + \Gamma_R)^2/4}
\label{Tr}
\end{equation}

 The result~(\ref{SLLL}, \ref{Tr}) is more general, than Exp.~(\ref{SME}),
obtained by means of master equation. 
If electrons do not interact, the Exp.~(\ref{SLLL})
is valid for any temperature. It can be simplified in the regime
$k_B T\gg \hbar\Gamma$. As we will show, in this limit the general result (\ref{SLLL})
coincides with  the result (\ref{SME}) of master equation.
It is easier to perform the calculation if one first evaluates the $\chi$-dependent
current $I(\chi) = (ie/ t_0)\partial S/\partial\chi$. It reads
\begin{gather}
\label{Ichi}
 I(\chi) = \frac{1}{2\pi} \int d\epsilon\, 
\left[ f_{(+)}(\epsilon) e^{i\chi} - f_{(-)}(\epsilon) e^{-i\chi}  \right]
 \times  \\ 
\Bigl\{T^{-1}_r(\epsilon) +  
\left[ f_{(-)}(\epsilon)( e^{-i\chi} -1) + f_{(+)}(\epsilon)( e^{i\chi} -1) \right]  
\nonumber
\Bigr\} 
\end{gather}
 Let us consider the situation when the resonant level is placed between the chemical
potentials $\mu_{L\{R\}}$ in the leads. 
Since we assumed that $k_B T\gg \Gamma_{L(R)}$, the main contribution
comes from the Lorentz peak
and one can put $\epsilon=\epsilon_i$ in the Fermi functions. 
Therefore we left only with the two poles 
$\epsilon_{1(2)} =\epsilon_i \pm i \sqrt{\cal{D}(\chi)}/2$ 
under the integrand~(\ref{Ichi}). Closing the integration contour in the upper or
lower half-plane we arrive at 
\begin{equation*}
I(\chi) = e\,\Gamma_L\Gamma_R
\left[ f_{(+)}(\epsilon_i)e^{i\chi} - f_{(-)}(\epsilon_i) e^{-i\chi}\right]/\sqrt{\cal D(\chi)}
\end{equation*}
Integrating it over $\chi$ one finds for the 
$S(\chi) = (t_0/ie)\int_0^{\chi}I(\chi')d\chi'$ the result~(\ref{SME}) obtained
by means of master equation. 

Thus, we have verified the correspondence between 
two approaches to statistics in the non-interacting regime.
We have shown that one can reproduce the statistics~(\ref{SME})
on substituting $T_r(\epsilon)$ into the Exp.~(\ref{SLLL}) and assuming the regime
$k_B T\gg \hbar\Gamma$. As it was discussed previously, this is the condition, when the master
equation approach, and hence its consequence~(\ref{SME}), are valid.

\end{section}

\begin{section}{Summary}
  We have reviewed here a constructive theory of counting statistics for 
electron transfer in multi-terminal mesoscopic systems. We have covered two
opposite limit of weak and strong interaction.
For the case of weakly interacting electrons, when the conductance of the
system $G\gg G_Q$, the theory of FCS reduces to 
a circuit theory of $2\times 2$ matrices associated with Keldysh Green functions.  
In the Coulomb blockade limit, $G\ll G_Q$, the FCS methods turns out to be an extension of 
the usual master equation approach. 
We have applied these methods to study the FCS of charge transfer through 
the three-terminal quantum dot. Surprisingly, the FCS has a similar qualitative
features both in weakly and strongly interacting regimes. 
We found that Coulomb interaction suppresses the 
relative probabilities of big current fluctuations in the dot. We
have also reviewed the scattering approach to FCS in multi-terminal circuits.
Then by considering the generic model of a single resonance level, we 
have established the equivalence of scattering and master equation approaches to FCS.

   The theories presented enables one for easy theoretical prediction of the FCS
for a given practical layout. Thereby they facilitate experimental 
activities in this direction. Up to now, only the noise has been measured.
In our opinion, the measurements of FCS can be easily 
performed with {\it threshold detectors} that
produce a signal provided the current in a certain terminal exceeds the 
threshold value. If the threshold value exceeds the average current, the 
detector will be switched by this relatively improbable fluctuation of the current.
The signal rate will be thus proportional to the probability of these
fluctuations $P(I)$, the value given by the theory of FCS. 

\end{section}


\begin{thebibliography}{99}

\bibitem{BlanterReview} 
Ya. M. Blanter and M. B\"{u}ttiker, Phys. Rep. {\bf 336}, 1 (2000).

\bibitem{Frac_Charge} L. Saminadayar, D. C. Glattli, Y. Jin, and B. Etienne, 
Phys. Rev. Lett. {\bf 79}, 2526 (1997). 
; R. de-Picciotto, M. Reznikov, M. Heiblum, V. Umansky, G. Bunin, and D. Mahalu, 
Nature {\bf 389}, 162 (1997).

\bibitem{Cron}  R. Cron, M. F. Goffman, D. Esteve, and C. Urbina, Phys. Rev. 
Lett. {\bf 86}, 4104 (2001). 

\bibitem{Chaotic} S.\ Oberholzer, E.\ V.\ Sukhorukov, C.\ Strunk, C.\ Sch\"onenberger.
T.\ Heinzel, and  M. Holland, Phys. Rev. Lett. {\bf 86}, 2114 (2001)

\bibitem{Kozhe}  A. A. Kozhevnikov, R. J. Schoelkopf, and D. E. Prober, Phys. 
Rev. Lett  {\bf 84}, 3398 (2000).

\bibitem{Jehl} 
X.\ Jehl {\it et.~al.}, Nature (London) {\bf 405}, 50 (2000).

\bibitem{Buttiker}  M. B\"{u}ttiker, Phys. Rev. B {\bf 46}, 12485 (1992).

\bibitem{Tarucha}  R. C. Liu, B. Odom, Y. Yamamoto, and S. Tarucha, Nature 
{\bf 391}, 263 (1998)

\bibitem{Oliver} W.\ D.\ Oliver, J.\ Kim, R.\ C.\ Liu, Y.\ Yamamoto,
Science, {\bf 284}, 299 (1999)

\bibitem{Hall} M.\ Henny, S.\ Oberholzer, C.\ Strunk, T.\ Heizel, K.\ Ensslin, M.\ Holland,
C. Sch\"onenberger, Science {\bf 284}, 296 (1999).

\bibitem{LL}  L. S. Levitov and G. B. Lesovik, JETP Lett. {\bf 58}, 230 (1993).

\bibitem{LLL} L. S. Levitov, H.-W. Lee, and G. B. Lesovik,
Journal of Mathematical Physics, {\bf 37} (1996) 10.

\bibitem{Yakovets} H.\ Lee, L.\ S.\ Levitov, A.\ Yu.\ Yakovets, Phys. Rev. B, 
{\bf 51}, 4079 (1995)

\bibitem{BlanterSchomerus} Ya. M. Blanter, H. Schomerus, and C.W.J.
Beenakker, Physica E {\bf 11}, 1 (2001).

\bibitem{Andreev} A.\ Andreev and A.\ Kamenev, Phys. Rev. Lett., {\bf 85}, 1294 (2000)

\bibitem{Levitov1} L.\ S.\ Levitov, arXiv: cond-mat/0103617, see also the contribution to the 
present book

\bibitem{Mirlin} Y. Makhlin and A. D. Mirlin,  Phys. Rev. Lett., {\bf 87}, 276803 (2001) 


\bibitem{RefYuli} Yu. V. Nazarov, 
  Ann. Phys. (Leipzig) {\bf 8} Spec. Issue, SI-193 (1999), cond-mat/9908143.

\bibitem{General} Yu. V. Nazarov, {\it Generalized Ohm's Law}, in: 
Quantum Dynamics of Submicron Structures, eds. H. Cerdeira, B. Kramer, G. 
Schoen, Kluwer, 1995, p. 687.

\bibitem{Belzig} W.\ Belzig and Yu.\ V.\ Nazarov, Phys. Rev. Lett., {\bf 87}, 067006 (2001);
W.\ Belzig and Yu.\ V.\ Nazarov, Phys. Rev. Lett., {\bf 87}, 197006 (2001).

\bibitem{BlanterSukhor} Ya. M. Blanter, E. V. Sukhorukov, Phys. Rev.
Lett. {\bf 84}, 1280 (2000).

\bibitem{Andreev1} A. V. Andreev and E. G. Mishchenko 
Phys. Rev. B 64, 233316 (2001)  

\bibitem{Fazio} M.-S. Choi, F. Plastina, and R. Fazio 
Phys. Rev. Lett. {\bf 87}, 116601 (2001)  

\bibitem{NazBag} Yu. V. Nazarov, D. A. Bagrets, Phys. Rev. Lett. 88, 196801 (2002) 

\bibitem{Rammer}  J. Rammer and H. Smith, Rev. Mod. Phys. {\bf 58}, 323
(1986).

\bibitem{Larkin} A. I. Larkin and Yu. V. Ovchinninkov,
Sov. Phys. JETP {\bf 41}, 960 (1975); Sov. Phys. JETP {\bf 46}, 155 (1977).

\bibitem{Noneq} Yu. V.  Nazarov, Superlattices Microst.\ {\bf 25}, 1221 (1999). 

\bibitem{Aleiner} O.~Agam, I.~Aleiner and A.~Larkin, Phys.~Rev.~Lett., 
{\bf 85}, 3153 (2000).  

\bibitem{Langen} S. A. van Langen,  M. B\"{u}ttiker, Phys. Rev. B., 
{\bf 56}, R1680 (1997)

\bibitem{Beenakker} C.W.J. Beenakker, Rev. Mod. Phys., {\bf 69},  731, (1997)

\bibitem{vanKampen} 
N.G. van Kampen, {\it Stochastic processes in physcics and chemistry}, Rev. and enl. eddition, 
{Elsevier Scinece Publishes B.V., North-Holland, 1992}

\bibitem{IngoldNaz} 
 G.-L.\ Ingold, Yu.\ V.\  Nazarov, in {\it Single Charge Tunneling}, 
 NATO ASI Series B: {\bf 294}, ed. H.\ Grabert, M.\ H.\ Devoret (NewYork, 1992) 



\end{thebibliography}
\end{document}